\documentclass[12pt]{article}
\usepackage[dvips]{graphicx}
\usepackage{float}
\usepackage{epsfig}
\usepackage{ulem}
\usepackage{latexsym,amsmath,amsfonts,amssymb}
\usepackage[latin1]{inputenc}
\usepackage{rotating}
\usepackage[american]{babel}
\usepackage[dvips]{graphicx}
\usepackage{bbm}
\usepackage{color}
\usepackage{slashed}
\usepackage[unicode]{hyperref}
\usepackage{lscape}
\usepackage{bigints}
\def\im{Invent. Math.}

\def\hat{\widehat}
\def\a{\alpha}
\def\b{\beta}
\def\c{\gamma}
\def\d{\delta}
\def\f{\phi}               
\def\vf{\varphi}  
\def\tvf{\tilde{\varphi}}
\def\vp{\varphi}
\def\g{\gamma}
\def\h{\eta}
\def\j{\psi}
\def\k{\kappa}                    
\def\l{\lambda}
\def\m{\mu}
\def\n{\nu}
\def\o{\omega}  \def\w{\omega}

\def\q{\theta}  \def\th{\theta}                  
\def\r{\rho}                                     
\def\s{\sigma}                                   
\def\t{\tau}
\def\u{\upsilon}
\def\x{\xi}
\def\z{\zeta}
\def\pt{\tilde{\varphi}}
\def\tt{\tilde{\theta}}
\def\lab{\label}
\def\6{\partial}
\def\wg{\wedge}
\def\bpsi{\bar{\psi}}
\def\bt{\bar{\theta}}
\def\bvf{\bar{\varphi}}

\DeclareMathOperator{\tr}{tr}

\newcommand{\be}{\begin{equation}}
\newcommand{\ee}{\end{equation}}
\newcommand{\beq}{\begin{equation}}
\newcommand{\eeq}{\end{equation}}
\newcommand{\bea}{\begin{eqnarray}}
\newcommand{\eea}{\end{eqnarray}}
\newcommand{\nn}{\nonumber}

\newcommand{\ba}{\begin{eqnarray}}
\newcommand{\ea}{\end{eqnarray}}

\newcommand{\beqs}{\begin{eqnarray}}
\newcommand{\eeqs}{\end{eqnarray}}
\newcommand{\bal}{\begin{aligned}}
\newcommand{\eal}{\end{aligned}}
\makeatletter
\newcommand\setItemnumber[1]{\setcounter{enum\romannumeral\@enumdepth}{\numexpr#1-1\relax}}
\makeatother

\begin{document}
\baselineskip=15.5pt
\pagestyle{plain}
\setcounter{page}{1}

\def\del{{\partial}}
\def\vev#1{\left\langle #1 \right\rangle}
\def\cn{{\cal N}}
\def\co{{\cal O}}


\def\IC{{\mathbb C}}
\def\IR{{\mathbb R}}
\def\IZ{{\mathbb Z}}
\def\RP{{\bf RP}}
\def\CP{{\bf CP}}
\def\Poincare{{Poincar\'e }}
\def\tr{{\rm tr}}
\def\tp{{\tilde \Phi}}

\def\TL{\hfil$\displaystyle{##}$}
\def\TR{$\displaystyle{{}##}$\hfil}
\def\TC{\hfil$\displaystyle{##}$\hfil}
\def\TT{\hbox{##}}
\def\HLINE{\noalign{\vskip1\jot}\hline\noalign{\vskip1\jot}}
\def\seqalign#1#2{\vcenter{\openup1\jot
   \halign{\strut #1\cr #2 \cr}}}
\def\lbldef#1#2{\expandafter\gdef\csname #1\endcsname {#2}}
\def\eqn#1#2{\lbldef{#1}{(\ref{#1})}%
\begin{equation} #2 \label{#1} \end{equation}}
\def\eqalign#1{\vcenter{\openup1\jot
     \halign{\strut\span\TL & \span\TR\cr #1 \cr
    }}}

\def\eno#1{(\ref{#1})}
\def\href#1#2{#2}
\def\half{\frac{1}{2}}



\def\ads{{\it AdS}}
\def\adsp{{\it AdS}$_{p+2}$}
\def\cft{{\it CFT}}

\newcommand{\ber}{\begin{eqnarray}}
\newcommand{\eer}{\end{eqnarray}}

\newcommand{\beqar}{\begin{eqnarray}}
\newcommand{\cN}{{\cal N}}
\newcommand{\cO}{{\cal O}}
\newcommand{\cA}{{\cal A}}
\newcommand{\cT}{{\cal T}}
\newcommand{\cF}{{\cal F}}
\newcommand{\cC}{{\cal C}}
\newcommand{\cR}{{\cal R}}
\newcommand{\cW}{{\cal W}}
\newcommand{\eeqar}{\end{eqnarray}}
\newcommand{\tht}{\thteta}
\newcommand{\lm}{\lambda}\newcommand{\Lm}{\Lambda}


\newcommand{\nonu}{\nonumber}
\newcommand{\oh}{\displaystyle{\frac{1}{2}}}
\newcommand{\dsl}
   {\kern.06em\hbox{\raise.15ex\hbox{$/$}\kern-.56em\hbox{$\partial$}}}
\newcommand{\id}{i\!\!\not\!\partial}
\newcommand{\as}{\not\!\! A}
\newcommand{\ps}{\not\! p}
\newcommand{\ks}{\not\! k}
\newcommand{\D}{{\cal{D}}}
\newcommand{\dv}{d^2x}
\newcommand{\Z}{{\cal Z}}
\newcommand{\N}{{\cal N}}
\newcommand{\Dsl}{\not\!\! D}
\newcommand{\Bsl}{\not\!\! B}
\newcommand{\Psl}{\not\!\! P}

\newcommand{\eeqarr}{\end{eqnarray}}
\newcommand{\ZZ}{{\rm \kern 0.275em Z \kern -0.92em Z}\;}


\def\del{{\delta^{\hbox{\sevenrm B}}}} \def\ex{{\hbox{\rm e}}}
\def\azb{A_{\bar z}} \def\az{A_z} \def\bzb{B_{\bar z}} \def\bz{B_z}
\def\czb{C_{\bar z}} \def\cz{C_z} \def\dzb{D_{\bar z}} \def\dz{D_z}
\def\im{{\hbox{\rm Im}}} \def\mod{{\hbox{\rm mod}}} \def\tr{{\hbox{\rm Tr}}}
\def\ch{{\hbox{\rm ch}}} \def\imp{{\hbox{\sevenrm Im}}}
\def\trp{{\hbox{\sevenrm Tr}}} \def\vol{{\hbox{\rm Vol}}}
\def\rl{\Lambda_{\hbox{\sevenrm R}}} \def\wl{\Lambda_{\hbox{\sevenrm W}}}
\def\fc{{\cal F}_{k+\cox}} \def\vev{vacuum expectation value}
\def\nodiv{\mid{\hbox{\hskip-7.8pt/}}}
\def\ie{{\em i.e.}}
\def\ie{\hbox{\it i.e.}}

\def\CC{{\mathchoice
{\rm C\mkern-8mu\vrule height1.45ex depth-.05ex
width.05em\mkern9mu\kern-.05em}
{\rm C\mkern-8mu\vrule height1.45ex depth-.05ex
width.05em\mkern9mu\kern-.05em}
{\rm C\mkern-8mu\vrule height1ex depth-.07ex
width.035em\mkern9mu\kern-.035em}
{\rm C\mkern-8mu\vrule height.65ex depth-.1ex
width.025em\mkern8mu\kern-.025em}}}

\def\RR{{\rm I\kern-1.6pt {\rm R}}}
\def\NN{{\rm I\!N}}
\def\ZZ{{\rm Z}\kern-3.8pt {\rm Z} \kern2pt}
\def\IB{\relax{\rm I\kern-.18em B}}
\def\ID{\relax{\rm I\kern-.18em D}}
\def\II{\relax{\rm I\kern-.18em I}}
\def\IP{\relax{\rm I\kern-.18em P}}
\newcommand{\CS}{{\scriptstyle {\rm CS}}}
\newcommand{\CSs}{{\scriptscriptstyle {\rm CS}}}
\newcommand{\rc}{\nonumber\\}
\newcommand{\bear}{\begin{eqnarray}}
\newcommand{\eear}{\end{eqnarray}}

\newcommand{\LL}{{\cal L}}

\def\mani{{\cal M}}
\def\calo{{\cal O}}
\def\calb{{\cal B}}
\def\calw{{\cal W}}
\def\calz{{\cal Z}}
\def\cald{{\cal D}}
\def\calc{{\cal C}}

\def\to{\rightarrow}
\def\ele{{\hbox{\sevenrm L}}}
\def\ere{{\hbox{\sevenrm R}}}
\def\zb{{\bar z}}
\def\wb{{\bar w}}
\def\nodiv{\mid{\hbox{\hskip-7.8pt/}}}
\def\menos{\hbox{\hskip-2.9pt}}
\def\dr{\dot R_}
\def\drr{\dot r_}
\def\ds{\dot s_}
\def\da{\dot A_}
\def\dga{\dot \gamma_}
\def\ga{\gamma_}
\def\dal{\dot\alpha_}
\def\al{\alpha_}
\def\cl{{closed}}
\def\cls{{closing}}
\def\vev{vacuum expectation value}
\def\tr{{\rm Tr}}
\def\to{\rightarrow}
\def\too{\longrightarrow}


\def\a{\alpha}
\def\b{\beta}
\def\c{\gamma}
\def\d{\delta}
\def\e{\epsilon}           
\def\F{\Phi}
\def\f{\phi}               
\def\vf{\varphi}  \def\tvf{\tilde{\varphi}}
\def\vp{\varphi}
\def\g{\gamma}
\def\h{\eta}
\def\j{\psi}
\def\k{\kappa}                    
\def\l{\lambda}
\def\m{\mu}
\def\n{\nu}
\def\o{\omega}  \def\w{\omega}
\def\q{\theta}  \def\th{\theta}                  
\def\r{\rho}                                     
\def\s{\sigma}                                   
\def\t{\tau}
\def\u{\upsilon}
\def\x{\xi}
\def\X{\Xi}
\def\z{\zeta}
\def\pt{\tilde{\varphi}}
\def\tt{\tilde{\theta}}
\def\lab{\label}
\def\6{\partial}
\def\wg{\wedge}
\def\atanh{{\rm arctanh}}
\def\bpsi{\bar{\psi}}
\def\bt{\bar{\theta}}
\def\bvf{\bar{\varphi}}

%



\newfont{\namefont}{cmr10}
\newfont{\addfont}{cmti7 scaled 1440}
\newfont{\boldmathfont}{cmbx10}
\newfont{\headfontb}{cmbx10 scaled 1728}





\newcommand{\re}{\,\mathbb{R}\mbox{e}\,}
\newcommand{\hyph}[1]{$#1$\nobreakdash-\hspace{0pt}}
\providecommand{\abs}[1]{\lvert#1\rvert}
\newcommand{\Nugual}[1]{$\mathcal{N}= #1 $}
\newcommand{\sub}[2]{#1_\text{#2}}
\newcommand{\partfrac}[2]{\frac{\partial #1}{\partial #2}}
\newcommand{\bsp}[1]{\begin{equation} \begin{split} #1 \end{split} \end{equation}}
\newcommand{\calF}{\mathcal{F}}
\newcommand{\calO}{\mathcal{O}}
\newcommand{\calM}{\mathcal{M}}
\newcommand{\calV}{\mathcal{V}}
\newcommand{\bbZ}{\mathbb{Z}}
\newcommand{\bbC}{\mathbb{C}}
\newcommand{\cK}{{\cal K}}

\newcommand{\Thq}{\Theta\left(\r-\r_q\right)}
\newcommand{\Dq}{\d\left(\r-\r_q\right)}
\newcommand{\kten}{\kappa^2_{\left(10\right)}}
\newcommand{\pbi}[1]{\imath^*\left(#1\right)}
\newcommand{\ho}{\hat{\omega}}
\newcommand{\tth}{\tilde{\th}}
\newcommand{\tf}{\tilde{\f}}
\newcommand{\tj}{\tilde{\j}}
\newcommand{\tw}{\tilde{\omega}}
\newcommand{\tz}{\tilde{z}}
\newcommand{\prj}[2]{(\partial_r{#1})(\partial_{\j}{#2})-(\partial_r{#2})(\partial_{\j}{#1})}
\def\atanh{{\rm arctanh}}
\def\sech{{\rm sech}}
\def\csch{{\rm csch}}
\allowdisplaybreaks[1]

\def\red{\textcolor[rgb]{0.98,0.00,0.00}}

\newcommand{\Dan}[1] {{\textcolor{blue}{#1}}}

\numberwithin{equation}{section}

\newcommand{\Tr}{\mbox{Tr}}    


%

\setcounter{footnote}{0}
\renewcommand{\theequation}{{\rm\thesection.\arabic{equation}}}

\begin{titlepage}

\begin{center}

\vskip .5in 
\noindent

{\Large \bf{ New AdS$_2$ backgrounds and ${\cal N}=4$ Conformal Quantum Mechanics} }
\bigskip\medskip

Yolanda Lozano$^{a,}$\footnote{ylozano@uniovi.es}, Carlos Nunez$^{b,}$\footnote{c.nunez@swansea.ac.uk}, Anayeli Ramirez$^{a,}$\footnote{anayelam@gmail.com} and Stefano Speziali$^{b,}$\footnote{stefano.speziali6@gmail.com}\\

\bigskip\medskip
{\small

 $a$: Department of Physics, University of Oviedo,
Avda. Federico Garcia Lorca s/n, 33007 Oviedo, Spain
\vskip 3mm
 $b$: Department of Physics, Swansea University, Swansea SA2 8PP, United Kingdom}

\vskip .5cm 
\vskip .9cm 
     	{\bf Abstract }\vskip .1in
\end{center}

\noindent
We present a new infinite family of Type IIB  backgrounds with an AdS$_2$ factor, preserving ${\cal N}=4$ SUSY. For each member of the family we propose a precise dual Super Conformal Quantum Mechanics (SCQM). We provide holographic expressions for the number of vacua (the ``central charge''), Chern-Simons terms and other  non-perturbative aspects of the SCQM. We relate the ``central charge'' of
	the one-dimensional system with a combination of electric and magnetic fluxes in Type IIB. The Ramond-Ramond fluxes are used to propose  an extremisation principle for the central charge. Other physical and geometrical aspects of these conformal quantum mechanics are analysed.
\noindent
\vskip .5cm
\vskip .5cm
\vfill
\eject

\end{titlepage}

\setcounter{footnote}{0}

\tableofcontents
\newpage
\renewcommand{\theequation}{{\rm\thesection.\arabic{equation}}}
\section{Introduction and general idea}

A major line of research  motivated by  the Maldacena conjecture \cite{Maldacena:1997re} is the study of supersymmetric and conformal field theories in diverse dimensions. 
Since the early 2000's efforts have been dedicated to  the classification of Type II or M-theory backgrounds with AdS$_{d+1}$ factors.
  These backgrounds are conjecturally dual to SCFTs in $d$ dimensions with different amounts of SUSY. For the case in which the solutions are half-maximal supersymmetric, important progress in classifying string backgrounds and the mapping to quantum field theories has been achieved.

Indeed, for ${\cal N}=2$ SCFTs in four dimensions, the field theories studied in \cite{Gaiotto:2009we} have holographic duals first discussed in \cite{Gaiotto:2009gz}, and further elaborated (among other works) in \cite{ReidEdwards:2010qs}-\cite{Bah:2019jts}.  The case of five dimensional SCFTs was analysed from the field theoretical and holographic viewpoints in \cite{Brandhuber:1999np}-\cite{Uhlemann:2019ypp}, among many other interesting works.   An infinite family of six-dimensional ${\cal N}=(1,0)$ SCFTs was discussed  both from the field theoretical and holographic points of view in \cite{Apruzzi:2015wna}-\cite{Hanany:1997gh}. For three-dimensional  ${\cal N}=4$ SCFTs, the field theoretical aspects presented in \cite{Gaiotto:2008ak} were discussed holographically in \cite{DHoker:2007hhe}-\cite{Lozano:2016wrs}, among other works.
The case of  two-dimensional SCFTs  and their AdS$_3$ duals  is very attractive, not only  for the rich landscape of two dimensional CFTs, but also for the connection with the Physics of black holes. In this case, recent progress was reported for half-maximal supersymmetric (for AdS$_3$) backgrounds, see for example  \cite{Lozano:2015bra}-\cite{Filippas:2020qku}.
All these solutions  geometrise various perturbative and non-perturbative aspects of conformal field theories in diverse dimensions.

A natural extension is the study of backgrounds with an AdS$_2$ factor \cite{Cvetic:2000cj}-\cite{Corbino:2020lzq}.
These should be  dual to superconformal quantum mechanics (SCQM). The similarities between the superconformal algebras in one and two dimensions or, by duality, the geometric relations between AdS$_2$ and AdS$_3$-spaces, suggest in particular that the studies of  \cite{Lozano:2019emq}-\cite{Filippas:2020qku} could be extended to the AdS$_2$ case. Some studies involving AdS$_2$ geometries were motivated by developments in black holes Physics, whilst others drew inspiration  from a purely geometric or 
field theoretical viewpoint, or both \cite{Strominger:1998yg}-\cite{Aniceto:2020saj}.

Surprisingly, the case of AdS$_2$/CFT$_1$ is less understood than its higher dimensional cousins. Indeed, various subtleties take place in the study of AdS$_2$ backgrounds \cite{Maldacena:1998uz}-\cite{Harlow:2018tqv}. Let us summarise some of them.

 A conformal quantum mechanical theory needs to have only SL$(2, \mathbb{R})$ global symmetry (aside from possible supersymmetry and associated R-symmetry).
Nevertheless, the analysis of  \cite{Hartman:2008dq,Alishahiha:2008tv,Cadoni:1999ja},  
implies that whilst the isometry of AdS$_2$ is SL$(2, \mathbb{R})$, asymptotically the group of symmetry is one-copy of the Virasoro algebra.
 The central charge of the algebra is proportional to the inverse Newton's constant in two dimensions. 

The connection between AdS$_3$ and AdS$_2$ geometries was discussed from the field theory perspective  in \cite{Balasubramanian:2009bg,Azeyanagi:2007bj}. These authors prove that quantising a two dimensional CFT using Discrete Light Cone Quantisation (DLCQ) is equivalent to decoupling one of the chiral sectors of a CFT.  
  In this paper we  use these ideas to connect AdS$_3$ and AdS$_2$ string solutions in  geometrical fashion.

In the context of JT-gravity, the authors of   \cite{Cvetic:2016eiv} found  flows  interpolating between AdS$_3$  and AdS$_2$ spaces. These correspond to the reduction  of AdS$_3$ along a space-like direction. There may be a relation between those solutions at the IR fixed point,  and the backgrounds we find in this work.
%
%
%
The authors of \cite{Azeyanagi:2007bj} found that black holes with generic AdS$_2$ near horizon geometry have an entanglement entropy related to the two-dimensional Newton's constant, according to,
\begin{equation}
S_{EE}=\frac{1}{G_{N}^{(2) } } .\label{SEE2}\nonumber
\end{equation}
They show that this entanglement entropy coincides with the entropy of a black hole whose near horizon contains the AdS$_2$.
In the present paper we perform explicit holographic calculations that hint at a relation between three quantities: 
the number of vacuum states of the SCQM, the partition function for the one dimensional SCFT when formulated on a circle
and the entropy of a black hole that has AdS$_2$ near horizon geometry. 

We  enlarge  the classification of SCFTs and AdS$_2$-string backgrounds, dealing with the case of ${\cal N}=4$  SCQMs and AdS$_2$ string geometries with an SU$(2)$-isometry. This leads us to the study of 
SCQM that are more elaborated than those usually analysed in the bibliography.
We define our SCQM to be the strongly coupled IR fixed point of ${\cal N}=4$ UV-finite quantum mechanical  quiver theories, that we precisely specify. Our  new ${\cal N}=4$ AdS$_2$ background solutions in Type IIB,  are a trustable dual description of the CFT$_1$ dynamics, whenever the number of nodes of the quiver and the ranks of each gauge group are large. We also need that the flavour groups (geometrically realised on source-branes) are widely separated in the geometry (we refer to this as the flavour groups being ``sparse'').
 
We present precise proposals for the ${\cal N}=4$ conformal quantum mechanics.
We  study various aspects of the SCQMs using the dual backgrounds. These include: number of vacua, Chern Simons coefficients, symmetry breaking, expected values of Wilson lines, couplings, etc. We
uncover a novel and  intriguing relation between a suitably defined ``central charge'' (associated with the number of vacua above mentioned) and the product of electric and magnetic charges for each Type IIB  background. 

The contents of this work are distributed as follows. In Section \ref{geometria}  we review the AdS$_3$ backgrounds in massive IIA that act as ``seed'' for our new infinite family of AdS$_2$ solutions in Type IIB. 
We revisit the two-dimensional ${\cal N}=(0,4)$ SCFTs dual to these backgrounds and improve on the existing bibliography by discussing the superpotential terms. In Section \ref{A-type} we present our new family of AdS$_2$ backgrounds and study in detail various geometrical aspects. In Section \ref{FTH} we present a concrete proposal for our ${\cal N}=4$ SCQM and perform  holographic calculations that encode field theoretical aspects
of our strongly coupled CFT$_1$s, with some emphasis on the holographic central charge above mentioned. 

In Section \ref{HCCMP}  we discuss a connection between the number of vacua of the SCQM and the RR sector of our supergravity solutions. We show that the holographic central charge is related to a product of electric and magnetic charges of the D-branes present in the background. We also present a new extremal principle in supergravity  from which the central charge of the SCQM can be obtained.
Our results extend and generalise those in the existing literature by the inclusion of sources and boundaries. Moreover they suggest new ways for the construction of the extremising functionals. In Section \ref{closing} we present our conclusions, with an invitation to colleagues working on field theoretical aspects of ${\cal N}=4$ SCQM to check some of our predictions using their favourite exact methods. Various appendices complement geometrical aspects of the backgrounds. 
Field theoretical observables of the strongly coupled quantum mechanical system are also holographically computed.

\section{Seed backgrounds and associated CFTs}\label{geometria}
%
In this section we review discuss the solutions to massive IIA supergravity (with localised sources) obtained in the recent work \cite{Lozano:2019emq}. These backgrounds provide the ``seed'' from which the new AdS$_2$ supergravity solutions presented in this work are derived. New results will also be presented.

For brevity, we restrict ourselves to a particular case of the generic backgrounds in  \cite{Lozano:2019emq}. The generic case is analysed in Appendix \ref{appendixgeneral}.  The Neveu-Schwarz (NS) sector of these solutions reads,
\begin{align}
\mathrm{d} s^2&= \frac{u}{\sqrt{\hat{h}_4 h_8}}\bigg(\mathrm{d}s^2_{\text{AdS}_3}+\frac{h_8\hat{h}_4 }{4 h_8\hat{h}_4+(u')^2} \mathrm{d}s^2_{\text{S}^2}\bigg)+ \sqrt{\frac{\hat{h}_4}{h_8}} \mathrm{d} s^2_{\text{CY}_2}+ \frac{\sqrt{\hat{h}_4 h_8}}{u} \text{d}\rho^2,\label{kkktmba}\\
e^{-\Phi}&= \frac{h_8^{\frac{3}{4}} }{2\hat{h}_4^{\frac{1}{4}}\sqrt{u}}\sqrt{4h_8 \hat{h}_4+(u')^2},~~~~ H_3= \frac{1}{2} \mathrm{d} \left(-\rho+\frac{ u u'}{4 \hat{h}_4 h_8+ (u')^2}\right)\wedge\text{vol}_{\text{S}^2},\nn
\end{align}
where $\Phi$ is the dilaton, $H=\mathrm{d}B_2$ is the NS 3-form and the metric is written in string frame. The warping functions $\hat{h}_4$, $h_8$ and $u$ have support on $\rho$.
We denote $u'= \partial_{\rho}u$ and similarly for $\hat{h}_4', h_8'$. 
The RR fluxes are 
\begin{subequations}
\begin{align}
F_0&=h_8',\;\;\;F_2=-\frac{1}{2}\bigg(h_8- \frac{ h'_8 u'u}{4 h_8 \hat{h}_4+ (u')^2} \bigg)\text{vol}_{\text{S}^2},\label{eq:classIflux2}\\[2mm]
F_4&= -\bigg(\mathrm{d}\left(\frac{u u'}{2\hat{ h}_4}\right)+2 h_8  \text{d}\rho\bigg) \wedge\text{vol}_{\text{AdS}_3}- \partial_{\rho}\hat{h}_4\text{vol}_{\text{CY}_2}
,\label{eq:classIflux3}
\end{align}
\end{subequations}
with the higher fluxes related to them as $F_6=-\star_{10} F_4,~F_8=\star_{10} F_2,~F_{10}=-\star_{10} F_0$.
The background in  \eqref{kkktmba}-\eqref{eq:classIflux3}  is a SUSY solution  of the massive IIA equations of motion if the functions $\hat{h}_4,h_8,u$ satisfy (away from localised sources),
\begin{equation}
\hat{h}_4''(\rho)=0,\;\;\;\; h_8''(\rho)=0,\;\;\;\; u''(\rho)=0.\label{eqsmotion}
\end{equation}
The first two are Bianchi identities. Hence the presence of localised sources will be indicated by delta-function inhomogeneities. In contrast, $u''=0$ is a BPS equation.
 
The  Page fluxes, defined as  $\hat{F}=e^{-B_2}\wedge F$, are
 \begin{eqnarray}
\hat{F}_0&=&h_8',\;\;
\hat{F}_2=-\frac{1}{2}\bigg(h_8- h_8'(\rho-2\pi k)\bigg)\text{vol}_{\text{S}^2},\;\;\nn\\
\hat{F}_4&=&-\bigg(\partial_\rho\left(\frac{u u'}{2 \hat{h}_4}\right)+2 h_8\bigg)  \text{d}\rho \wedge\text{vol}_{\text{AdS}_3}
- \partial_{\rho}\hat{h}_4\text{vol}_{\text{CY}_2}.\label{eq:background}
\end{eqnarray}
We have allowed for large gauge transformations $B_2\to B_2 + { \pi k} \text{vol}_{\text{S}^2}$, for $k=0,1,...., P$. The transformations are performed every time we cross an interval $[2\pi k, 2\pi(k+1)]$. The $\rho$-direction begins at $\rho=0$ and ends at $\rho=2\pi(P+1)$. This will become apparent once the functions $\hat{h}_4, h_8, u$ are specified below.

%
%
Various particular solutions were analysed in \cite{Lozano:2019emq}. Here we consider an infinite family of backgrounds for which the $\hat{h}_4$, $h_8$ functions are piecewise continuous.  These were carefully studied in \cite{Lozano:2019zvg}-\cite{Lozano:2019ywa}, where a precise dual field theory was proposed.
%
 The above mentioned range  of the $\rho$-coordinate is determined by the vanishing of the functions $\hat{h}_4$ and $ h_8$.
Generically these functions read,
 \begin{equation} \label{profileh4sp}
\hat{h}_4(\rho)\!=\!\Upsilon\! \,h_4(\rho)\!=\!\!
                    \Upsilon\!\!\left\{ \begin{array}{ccrcl}
                       \frac{\beta_0 }{2\pi}
                       \rho & 0\leq \rho\leq 2\pi \\
                                     \alpha_k\! +\! \frac{\beta_k}{2\pi}(\rho-2\pi k) &~~ 2\pi k\leq \rho \leq 2\pi(k+1),\;\; k=1,..,P-1\\
                      \alpha_P-  \frac{\alpha_P}{2\pi}(\rho-2\pi P) & 2\pi P\leq \rho \leq 2\pi(P+1),
                                             \end{array}
\right.
\end{equation}
 \begin{equation} \label{profileh8sp}
h_8(\rho)
                    =\left\{ \begin{array}{ccrcl}
                       \frac{\nu_0 }{2\pi}
                       \rho & 0\leq \rho\leq 2\pi \\
                        \mu_k+ \frac{\nu_k}{2\pi}(\rho-2\pi k) &~~ 2\pi k\leq \rho \leq 2\pi(k+1),\;\;\;\; k:=1,....,P-1\\
                      \mu_P-  \frac{\mu_P}{2\pi}(\rho-2\pi P) & 2\pi P\leq \rho \leq 2\pi(P+1).
                                             \end{array}
\right.
\end{equation}
%
%
The quantities $(\alpha_k,\beta_k, \mu_k, \nu_k)$ are integration constants. By imposing continuity we determine,
\begin{equation}
\alpha_k=\sum_{j=0}^{k-1} \beta_j ,~~~\mu_k= \sum_{j=0}^{k-1}\nu_j.\label{definitionmupalphap}
\end{equation} 
Below, we summarise aspects of the two dimensional field theories dual to the backgrounds \eqref{kkktmba}-\eqref{eqsmotion}  for the solutions determined by eqs.(\ref{profileh4sp})-(\ref{profileh8sp}). We also present new aspects of these field theories.

\subsection{The associated dual SCFTs}\label{2dSCFT}
\begin{figure}[h!]
    \centering
    {{\includegraphics[width=10cm]{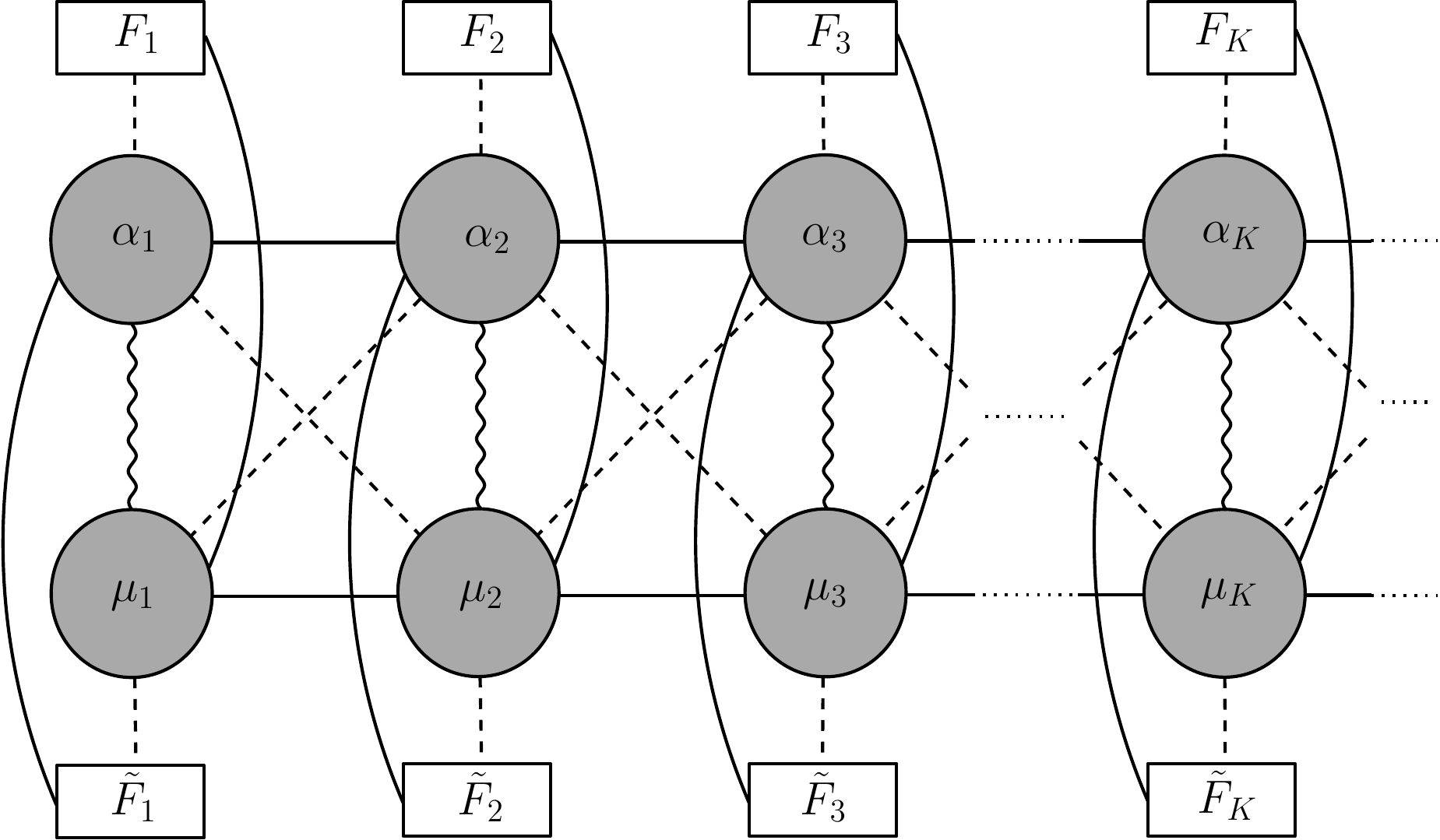} }}%

\caption{ A generic quiver field theory whose IR is dual to the holographic background defined by the functions in \eqref{profileh4sp}-\eqref{profileh8sp}. The solid black lines represent $(4,4)$ hypermultiplets, the wavy lines represent  $(0,4)$ hypermultiplets and  the dashed lines represent  $(0,2)$ Fermi multiplets. ${\cal N}=(4,4)$ vector multiplets are the degrees of freedom  in each gauged node.}
\label{figurageneral}

\end{figure}

As was explained in the papers  \cite{Lozano:2019zvg}-\cite{Lozano:2019ywa}, for the functions  $\hat{h}_4,h_8, u$ in eqs.(\ref{profileh4sp})-(\ref{profileh8sp}),  the backgrounds in eqs.\eqref{kkktmba}-\eqref{eqsmotion} 
are 
associated with a Hanany-Witten  \cite{Hanany:1996ie}  set-up indicated in Table \ref{D6-NS5-D8-D2-D4-first}.
 \begin{table}[ht]
	\begin{center}
		\begin{tabular}{| l | c | c | c | c| c | c| c | c| c | c |}
			\hline		    
			& 0 & 1 & 2 & 3 & 4 & 5 & 6 & 7 & 8 & 9 \\ \hline
			D2 & x & x & &  &  &  & x  &   &   &   \\ \hline
			D4 & x & x &  &  &  &   &  & x & x & x  \\ \hline
			D6 & x & x & x & x & x & x & x  &   &   &   \\ \hline
			D8 & x & x &x  & x & x &  x &  & x & x & x  \\ \hline
			NS5 & x & x &x  & x & x & x  &   &   &  &  \\ \hline
		\end{tabular} 
	\end{center}
	\caption{BPS brane intersection underlying the geometry in \eqref{kkktmba}-\eqref{eqsmotion}. The directions $(x^0,x^1)$ are the directions where the 2d dual CFT lives. The directions $(x^2, \dots, x^5)$ span the CY$_2$, on which the D6 and the D8-branes are wrapped. The coordinate $x^6$ is the direction associated with $\rho$. Finally $(x^7,x^8,x^9)$ are the transverse directions realising an SO(3)-symmetry associated with the isometries of S$^2$.}   
	\label{D6-NS5-D8-D2-D4-first}	
\end{table} 
Using this brane set-up, dual  two-dimensional  CFTs with ${\cal N}=(0,4)$ SUSY were proposed. These CFTs describe the low energy, strongly coupled dynamics of  two dimensional quantum field theories. The field theories are  encoded by the quiver in Figure \ref{figurageneral}.
The difference between this quiver and those proposed in  \cite{Lozano:2019zvg}-\cite{Lozano:2019ywa} is the presence of $(4,4)$ matter connecting flavour and colour groups. These correspond in Figure \ref{figurageneral} to the vertical (bent) lines.
In the limit that makes the holographic backgrounds trustable (that is, long quivers with large ranks and  sparse flavour groups), these (4,4) hypermultiplets do not affect the matching of observables discussed in  \cite{Lozano:2019zvg}-\cite{Lozano:2019ywa}. In fact, their contribution is subleading and not captured by supergravity.

The absence of gauge anomalies constrains the ranks of the flavour groups to be
\begin{equation}
F_k=\nu_{k-1}-\nu_{k},\;\;\;\; \tilde{F}_k=\beta_{k-1}-\beta_{k}.\label{flavours}
\end{equation}
These are precisely the quantised numbers of D8 and D4  flavour (source) branes derived from eq.(\ref{eq:background})
These conditions are unchanged by the presence of the ${\cal N}=(4,4)$ bifundamentals connecting flavour and colour groups, which  (being vectorial) do not count towards the anomaly. 

Numerous checks for the validity of this proposal were presented in  \cite{Lozano:2019zvg}-\cite{Lozano:2019ywa}.
The right-handed central charge of the SCFTs is computed by identifying it with the $U(1)_R$ 
current two-point function.
The works \cite{Tong:2014yna,Putrov:2015jpa} found that for a generic quiver with $n_{hyp}$ hypermultiplets and $n_{vec}$ vector multiplets the central charge is,
\begin{equation}
c_{CFT}=6 (n_{hyp}- n_{vec}).\label{centralcft}
\end{equation}
The papers \cite{Lozano:2019zvg}-\cite{Lozano:2019ywa} present a variety of examples of long linear quivers with sparse flavour groups and large ranks for each of the nodes. In each of these qualitatively different examples, it was found that the field theoretical central charge of eq.(\ref{centralcft})
coincides with the holographic central charge (at leading order, when the background is a trustable dual description to the CFT).  Note that this matching is not changed by the presence of the extra $(4,4)$ hypermutiplets mentioned above.  The ``sparse'' character of the flavour groups makes their contribution subleading.

The expression for the holographic central charge derived in \cite{Lozano:2019zvg}-\cite{Lozano:2019ywa} is,
\begin{equation}
c_{hol}= \frac{3\pi}{2 G_N}\text{Vol}_{\text{CY}_2}\int_0^{2\pi (P+1)} \hat{h}_4 h_8 \text{d}\rho= \frac{3}{\pi} \int_0^{2\pi (P+1)} {h}_4 h_8 \text{d}\rho  .\label{chol}
\end{equation}
We used that $G_N= 8\pi^6$ (with $g_s=\alpha'=1$) and that $\Upsilon \text{Vol}_{\text{CY}_2}=16\pi^4$.
\subsubsection{Superpotential}\label{secsupxx}
\begin{figure}[h!]
    \centering
    {{\includegraphics[width=8.2cm]{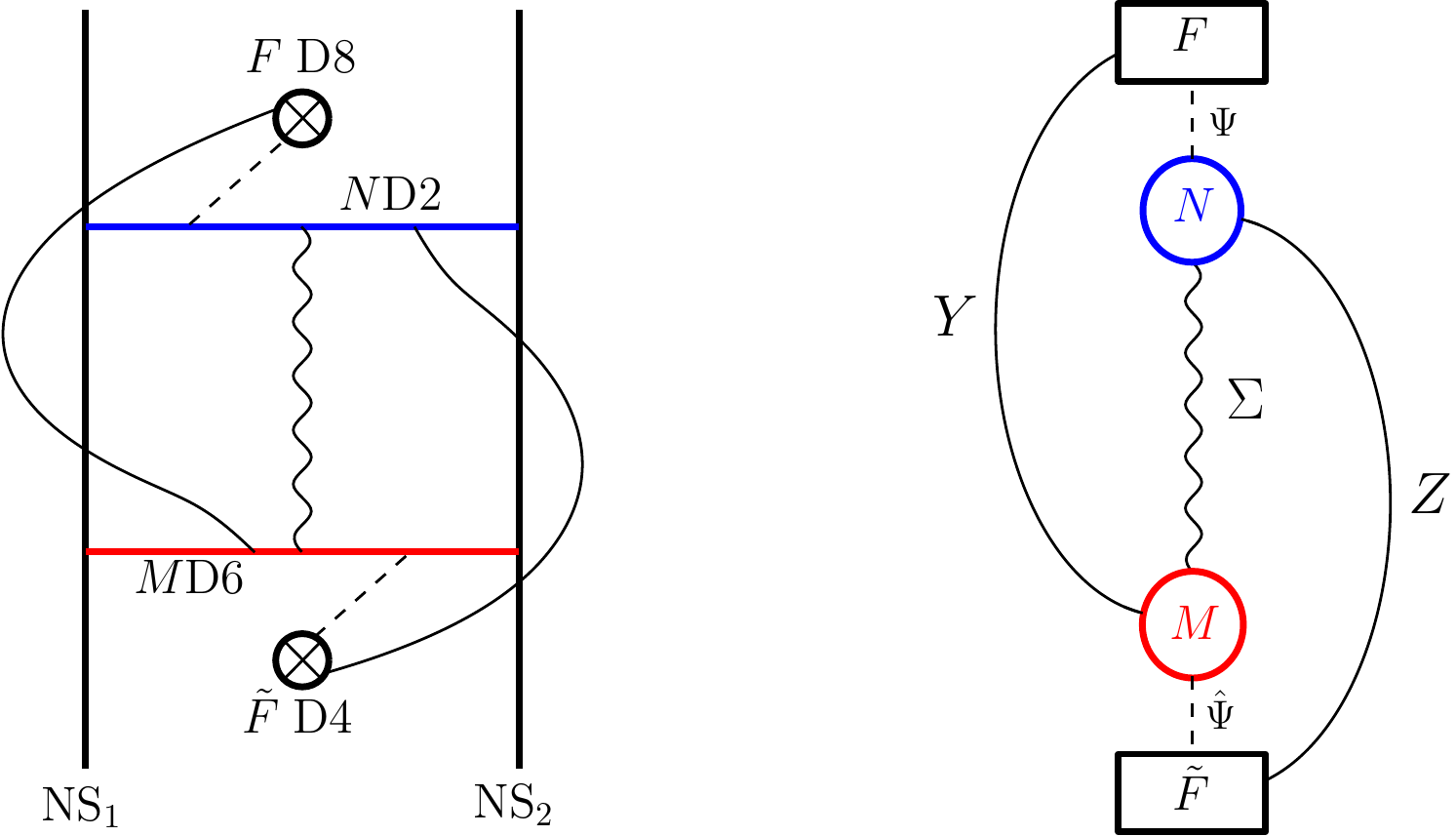} }}%
%
%
\caption{ On the left, we plot one cell in the Hanany-Witten set-up, in between the NS-five branes NS$_1$ and NS$_2$. The solid black lines represent $(4,4)$ hypermultiplets, the curvy lines $(0,4)$ hypermultiplets and  the dashed lines $(0,2)$ Fermi multiplets. On the right, we plot the field content of one cell in the quiver (same convention). $Y,Z$ are $(4,4)$ hypers, $\Sigma$ denotes a $(0,4)$ hyper. The $(0,2)$ Fermi fields are denoted by $(\Psi,\hat{\Psi})$. The gauge nodes contain $(4,4)$ vector multiplets. }
\label{LAG4}

\end{figure}

Now, we present a new development, adding value to this review-section. Let us  discuss  the  superpotential terms that can be written due to the presence of the $(4,4)$ hypermultiplets connecting D2-D4 and D6-D8 branes.
 In two dimensions  with ${\cal N}=(0,4)$ SUSY, we can  write interactions in terms of a superpotential  $W$ \cite{Tong:2014yna}-\cite{Witten:1993yc},
\begin{equation}
S= \int \mathrm{d}^2 x \mathrm{d}\theta^+ W,\;\;\;\;\; W= \Psi_a J^a(\Phi_i).\label{W-superpotential}
\end{equation}
Studying the strings stretched between the different branes in the Hanany-Witten set-up, we find the massless fields described in Figure \ref{LAG4} (left side). We depict only one ``interval'' of the whole  Hanany-Witten set-up. The blue and red lines denote $N$-D2 branes and $M$-D6 branes, there are also $F$ D8 and $\tilde{F}$ D4 flavour branes. The D2 and D6 colour branes are joined by a wavy line, representing a $(0,4)$ hyper (denoted by $\Sigma$ in the right figure). Dashed lines represent $(0,2)$ Fermi multiplets, joining D2-D8 and D4-D6 pairs. These are denoted by $\Psi,\hat{\Psi}$ in the right figure.  We also have solid black lines, representing $(4,4)$ hypermultiplets, joining D2-D4 and D6-D8 branes and denoted by $Y,Z$ on the right panel of Figure \ref{LAG4}. This ``interval'' is connected via $(0,2)$ Fermi multiplets and $(4,4)$ hypermultiplets with a similar next-interval as indicated in Figure \ref{figurageneral}. 

%
%
%

As discussed above, these new $(4,4)$ matter fields $Y,Z$ have no-effect on anomalies and their effect on the central charge is subleading.
Their presence was emphasised in 
\cite{Filippas:2020qku}. They allow to write a superpotential term.

%
%
The superpotential interaction is obtained by closing the ``triangle'', contracting indexes appropriately in the circuit D8-D2-D6-D8 and D4-D6-D2-D4. This suggests that we should include cubic superpotential terms of the form,
\begin{equation}
W\sim Y \Sigma \Psi + Z \Sigma \hat{\Psi}. \label{W-final}
\end{equation}

%
%
%
%
In Appendix \ref{appendix1} we give more details about the Lagrangian associated with the quiver QFT in Figure \ref{figurageneral}.
Putting together all this information, the full Lagrangian describing the UV dynamics is written there. This dynamics conjecturally flows in the IR to a CFT with small ${\cal N}=(0,4)$ SUSY  \cite{Lozano:2019zvg}-\cite{Lozano:2019ywa}.

 After this summary of the ``seed'' backgrounds and dual SCFTs, let us now focus on the new infinite family of backgrounds and the associated SCQMs.
\section{New Type IIB backgrounds}\label{A-type}
In this section we present a new infinite family of  AdS$_2$ backgrounds of Type IIB supergravity. They are obtained by applying T-duality on the seed backgrounds defined by  eqs.(\ref{kkktmba})-(\ref{eqsmotion}),  along a direction inside AdS$_3$. 

Consider the backgrounds of eqs.\eqref{kkktmba}-\eqref{eqsmotion}. Write  AdS$_3$ as a fibration over AdS$_2$,
\begin{eqnarray}
& &\text{d} s^2_{\text{AdS}_3} = \frac{1}{4}\left[\left(\text{d}\tilde{\psi} +\eta\right)^2 + \text{d} s^2_{\text{AdS}_2}\right]\qquad\text{with}\qquad \text{d} \eta= \text{vol}_{\text{AdS}_2}.\label{nara}\\
& & \text{d} s^2_{\text{AdS}_2}= -\mathrm{d} t^2\cosh^2 r +  \mathrm{d} r^2,\;\;\; \eta= -\sinh r \text{d}t.\;\;\; 
\nonumber
\end{eqnarray}

We T-dualise on the fibre direction to obtain the new solutions (more general configurations are discussed in Appendix \ref{appendixgeneral}). These backgrounds have the structure $\text{AdS}_2\times \text{S}^2\times \textrm{CY$_2$}\times \text{I}_\rho\times \text{S}^1_\psi$. The NS sector reads,
\begin{equation}\label{t dualised background NS}
\begin{split}
\text{d}s^2 &= \frac{u}{\sqrt{\hat{h}_4 h_8}} \left( \frac{1}{4}\text{d}s^2_{\text{AdS}_2} + \frac{\hat{h}_4 h_8}{4 \hat{h}_4 h_8 + (u')^2} \text{d}s^2_{\text{S}^2} \right) + \sqrt{\frac{\hat{h}_4}{h_8}} \text{d}s^2_{\text{CY}_2} + \frac{\sqrt{\hat{h}_4 h_8}}{u} (\text{d} \rho^2 +  \text{d} \psi^2 )\, , \\
e^{- 2 \Phi}&= \frac{h_8}{4\hat{h}_4} \big(4 \hat{h}_4 h_8 + (u')^2 \big) \, , \quad H_{3} = \frac{1}{2} \text{d} \bigg( - \rho + \frac{u u'}{4 \hat{h}_4 h_8 + (u')^2} \bigg) \wedge \text{vol}_{\text{S$^2$}} + \frac{1}{2}\text{vol}_{\text{AdS$_2$}} \wedge \text{d} \psi \, ,
\end{split}
\end{equation}
where $\psi$ is the T-dual-coordinate, with range $[0,2\pi]$.

The RR sector is given by
\begin{equation}\label{t dualised background RR}
\begin{split}
F_{1} &= h_8' \text{d} \psi \, , \quad F_{3} =  - \frac{1}{2} \Big( h_8 - \frac{h_8' u' u}{4 h_8 \hat{h}_4+(u')^2} \Big) \text{vol}_{\text{S$^2$}} \wedge \text{d} \psi + \frac{1}{4} \left( \text{d} \bigg( \frac{u'u}{2 \hat{h}_4} \bigg) + 2 h_8 \text{d} \rho \right) \wedge \text{vol}_{\text{AdS$_2$}} \, , \\
F_{5} &= -(1 + \star) \, \hat{h}_4' \, \text{vol}_{\text{CY$_2$}} \wedge \text{d} \psi = - \hat{h}_4' \, \text{vol}_{\text{CY$_2$}} \wedge \text{d} \psi + \frac{\hat{h}_4' h_8 u^2}{4 \hat{h}_4 (4 \hat{h}_4 h_8 + (u')^2)} \text{vol}_{\text{AdS$_2$}} \wedge \text{vol}_{\text{S$^2$}} \wedge \mathrm{d} \rho \, , \\
F_{7} & =  \frac{4 \hat{h}_4^2 h_8 -u u' \hat{h}_4' + \hat{h}_4  (u')^2}{8 \hat{h}_4 h_8 + 2 (u')^2} \text{vol}_{\text{CY$_2$}}  \wedge \text{vol}_{\text{S$^2$}} \wedge \text{d} \psi - \frac{4 \hat{h}_4 h_8^2 - u u' h_8' + h_8  (u')^2}{8 h_8^2} \text{vol}_{\text{AdS$_2$}} \wedge \text{vol}_{\text{CY$_2$}} \wedge \mathrm{d} \rho \, , \\
F_{9}&= -\frac{\hat{h}_4 h_8' u^2}{4 \hat{h}_8 (4 \hat{h}_4 h_8 + (u')^2)} \text{vol}_{\text{AdS$_2$}} \wedge\text{vol}_{\text{CY$_2$}} \wedge \text{vol}_{\text{S$^2$}} \wedge \mathrm{d} \rho \, ,
\end{split}
\end{equation}
where $F_{7}=- \star F_{3}=$ and $F_{9}=\star F_{1}$.
We also quote the explicit expression of $\star H_{3}$,
\begin{equation}
\star H_{3} = \frac{2 \hat{h}_4^2}{4 \hat{h}_4 h_8 + (u')^2} \text{vol}_{\text{CY$_2$}} \wedge \text{vol}_{\text{S$^2$}} \wedge  \mathrm{d} \rho - \frac{\hat{h}_4' h_8 u u' + \hat{h}_4 u' \left(u h_8' + h_8 u' \right) + 4 \hat{h}_4^2 h_8^2}{2h_8^2 \big(4 \hat{h}_4 h_8 + u'^2 \big)} \text{vol}_{\text{AdS$_2$}} \wedge \text{vol}_{\text{CY$_2$}} \wedge \mathrm{d} \psi \, \nonumber.
\end{equation}
One can check that the Type IIB equations of motion are satisfied
imposing the BPS equations and Bianchi identities: $u''=0$ and $\hat{h}_4''=0$, $h_8''=0$. A violation of the Bianchi identities is admissible at points where brane sources are located. We consider solutions like those in eqs.(\ref{profileh4sp})-(\ref{profileh8sp}).

\begin{figure}[h!]
    \centering
    {{\includegraphics[width=7.3cm]{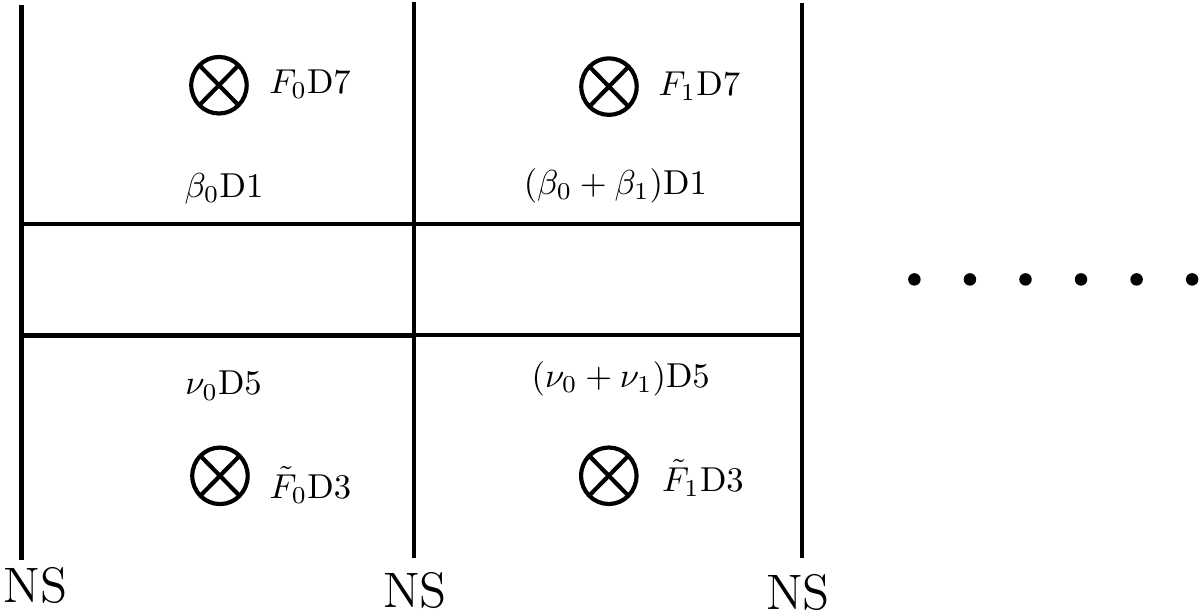} }}%

\caption{ The Hanany-Witten set-up corresponding to the background in eqs.(\ref{t dualised background NS})-(\ref{t dualised background RR}).}
\label{BraneSetup}
\end{figure}

We perform a large gauge transformation $B_2\to B_2+ k\pi  \text{vol}_{\text{S$^2$}}$. This naturally divides the interval $\text{I}_\rho$ in $(P+1)$-cells of size $2\pi$. The Page forms $\hat{F}= e^{-B_2}\wedge F$ are,
%
\begin{equation}\label{fluxes2}
\begin{split}
\hat{F}_{1} &= h'_8 \, \mathrm{d} \psi \, , \\
\hat{F}_{3} &= \frac{1}{2} \left( h'_8(\rho-2\pi k)  - h_8 \right) \, \text{vol}_{\text{S$^2$}} \wedge \mathrm{d} \psi  + \frac{1}{4} \bigg( \frac{u' \big( \hat{h}_4 u' - u \hat{h}_4' \big)}{2 \hat{h}_4^2} + 2 h_8 \bigg)\text{vol}_{\text{AdS$_2$}} \wedge \mathrm{d} \rho  \, , \\
\hat{F}_{5} &= \frac{1}{16} \bigg(  \frac{\left(u - (\rho-2\pi k)  u' \right) ( u  \hat{h}_4' - \hat{h}_4  u'  )}{ \hat{h}_4^2} + 4 (\rho-2\pi k)  h_8     \bigg) \text{vol}_{\text{AdS$_2$}} \wedge \text{vol}_{\text{S$^2$}} \wedge \mathrm{d} \rho - \hat{h}_4' \text{vol}_{\text{CY$_2$}} \wedge \mathrm{d} \psi \, , \\
\hat{F}_{7} &= \frac{1}{2} \left( \hat{h}_4 - (\rho-2\pi k) \hat{h}_4' \right)  \text{vol}_{\text{S$^2$}}  \wedge \text{vol}_{\text{CY$_2$}} \wedge \mathrm{d} \psi -\left(  \frac{4 \hat{h}_4 h_8^2 - u u' h_8' + h_8 (u')^2}{8 h_8^2} \right)\text{vol}_{\text{AdS$_2$}} \wedge \text{vol}_{\text{CY$_2$}} \wedge \mathrm{d} \rho  \, ,\\
\hat{F}_{9}&=- \left(\frac{ u^2 h_8' - h_8 u u'  +(\rho-2\pi k)( h_8 u'^2 - u u' h'_8 + 4\hat{h}_4 h_8^2 )}{16 h_8^2}\right)\text{vol}_{\text{AdS$_2$}} \wedge \text{vol}_{\text{S$^2$}}\wedge  \text{vol}_{\text{CY$_2$}}\wedge \text{d}\rho.
\end{split}
\end{equation}
To describe the brane set-up, we use that $\hat{h}_4$ and $h_8$ are continuous polygonal functions with discontinuous derivatives, as in  eqs.(\ref{profileh4sp})-(\ref{profileh8sp}). We compute,
\begin{eqnarray}
& & \mathrm{d}\hat{F}_{1}= h_8'' \text{d}\rho\wedge \text{d}\psi,\;\;\;
 \mathrm{d}\hat{F}_{3}= -\frac{1}{2}h_8'' \times(\rho-2\pi k) \mathrm{d}\rho \wedge  \text{vol}_{\text{S$^2$}} \wedge \mathrm{d} \psi ,\label{caxa}\\
& & \mathrm{d}\hat{F}_{5}= -\hat{h}_4'' \mathrm{d}\rho\wedge \text{vol}_{\text{CY$_2$}} \wedge \mathrm{d} \psi ,\;\;\mathrm{d}\hat{F}_{7}= -\frac{1}{2}\hat{h}_4'' \times(\rho-2\pi k) \mathrm{d}\rho \wedge \text{vol}_{\text{S$^2$}}  \wedge \text{vol}_{\text{CY$_2$}} \wedge \mathrm{d} \psi ,\nonumber\\
& & \mathrm{d}\hat{F}_{9}= 0,
\end{eqnarray}
with
\begin{eqnarray}
& & \hat{h}_4''=\frac{1}{2\pi}\sum_{j=1}^P (\beta_{j-1}-\beta_{j}) \delta(\rho-2\pi j),\;\;\;\; h_8''=\frac{1}{2\pi}\sum_{j=1}^P (\nu_{j-1}-\nu_{j} )\delta(\rho-2\pi j),\label{nana}\\
& & \hat{h}_4'' \times(\rho-2\pi k) = h_8'' \times(\rho-2\pi k) =x \delta(x)=0.\nonumber
\end{eqnarray}
Inspecting the Page fluxes, the electric parts tell us what type of branes we have in the system. 
The exterior derivative
of the dual magnetic form $\mathrm{d} \hat{F}_{8-p}$ being nonzero, indicates that the Dp brane is a source in the background (flavour branes). In contrast, $\mathrm{d}\hat{F}_{8-p}=0$
indicates that these branes are dissolved into fluxes (colour branes). We then find a brane set-up consisting of (colour) D1 and D5 branes, extending in between NS-five branes. This is complemented by (sources) D3 and D7 branes. There are also fundamental strings dissolved into flux.
%
We list the brane content in Table \ref{Table brane web set type IIA} and the associated Hanany-Witten set-up in Figure \ref{BraneSetup}. 
 \begin{table}[ht]
	\begin{center}
		\begin{tabular}{| l | c | c | c | c| c | c| c | c| c | c |}
			\hline		    
			& 0 & 1 & 2 & 3 & 4 & 5 & 6 & 7 & 8 & 9 \\ \hline
			D1 & x &  & &  &  & x &   &   &   &   \\ \hline
			D3 & x &  &  &  &  &   & x & x & x &   \\ \hline
			D5 & x & x & x & x & x & x &   &   &   &   \\ \hline
			D7 & x & x &x  & x & x &   &x  & x & x &   \\ \hline
			NS5 & x & x &x  & x & x &   &   &   &   & x  \\ \hline
			F1 &x& & & & & & & & & x \\ \hline
		\end{tabular} 
\caption{Brane set-up underlying the geometry in (\ref{t dualised background NS})-(\ref{t dualised background RR}). 
$x^0$ is the time direction of the ten dimensional spacetime. The directions $(x^1, \dots , x^4)$ span the CY$_2$, $x^5$ is the direction associated with $\rho$, $(x^6, x^7, x^8)$ are the transverse directions realising the SO(3)-symmetry of the $S^2$, and $x^9$ is the $\psi$ direction.}
\label{Table brane web set type IIA}
\end{center}
\end{table}

Let us study the quantised Page charges, defined by integrating the Page magnetic flux\footnote{The relevant constants are, 
\begin{equation}
T_{\text{Dp}} =\frac{1}{(2\pi)^{p} g_s \alpha'^{\frac{p+1}{2}}},\;\;\; 2\kappa_{10}^2= (2\pi)^7 g_s^2 \alpha'^4,\;\;\; T_{\text{NS5}}=\frac{1}{(2\pi)^5 g_s^2\alpha'^3},\;\;\; \alpha'=g_s=1.\nonumber
\end{equation}
}, 
\begin{equation}\label{Dp-Pagecharge}
Q_{\text{Dp}}=\frac{1}{2\kappa_{10}^2 T_{\text{Dp}}}\int \hat{F}_{8-p}=\frac{1}{(2\pi)^{7-p}}\int \hat{F}_{8-p}\, .
\end{equation}
The functions $\hat{h}_4, h_8$  are as those in eqs.(\ref{profileh4sp})-(\ref{profileh8sp}). The integrals over volumes are,
\begin{equation}
\Upsilon \text{Vol}_{\text{CY$_2$} }=16\pi^4,\;\;\;\; \int \text{d}\psi=\text{Vol}_{\psi } =2\pi,\;\;\;  \text{Vol}_{\text{S}^2 }=4\pi .\nonumber
\end{equation} 
The different brane charges in each interval $[2\pi k, 2\pi (k+1)]$ are
\begin{eqnarray}
& & Q_{\text{D1}}=\frac{1}{(2\pi)^6}\int_{\Sigma_7} \hat{F}_{7}= \left(\frac{ \Upsilon \text{Vol}_{\text{CY$_2$} }     }{16\pi^4} \right) \times \left(\frac{\text{Vol}_{\text{S}^2 }  }{4\pi} \right) \times \left(\frac{ \text{Vol}_{\psi } }{2\pi} \right) \left( h_4 -h_4'(\rho-2\pi k) \right)= \alpha_k,\nonumber\\
& & Q_{\text{D3}}=\frac{1}{16\pi^4}\int_{\Sigma_5} \hat{F}_{5}=  \frac{1}{16\pi^4}\int_{[\rho,\Sigma_5]} \mathrm{d}\hat{F}_{5}=\left(\frac{ \Upsilon \text{Vol}_{\text{CY$_2$} }     }{16\pi^4} \right) \times \text{Vol}_\psi~ \int \text{d}\rho h_4''=\beta_{k-1}-\beta_{k},\nonumber\\
& & Q_{\text{D5}}=\frac{1}{4\pi^2}\int_{\Sigma_3} \hat{F}_{3} =\left(\frac{  \text{Vol}_{\text{S$^2$}      }}{4\pi}\right)\times \left(\frac{  \text{Vol}_\psi }{2\pi} \right) \left(h_8-h_8'(\rho-2\pi k) \right)=\mu_k,\nonumber\\
& & Q_{\text{D7}}=\int_{\Sigma_1} F_{1}=\int_{[\rho,\Sigma_1]} \mathrm{d} F_{1}=  \text{Vol}_\psi ~\int h_8'' \text{d}\rho =\nu_{k-1}-\nu_{k}.\label{cargaspageA}
\end{eqnarray}
Notice that we have used the expression for the second derivatives in eq.(\ref{nana}).

The structure of singularities in the two ends of the $\rho$-interval is studied in Appendix \ref{rhointerval}. Referring to the set-up in Table \ref{Table brane web set type IIA} and Figure \ref{BraneSetup}, in the $[2\pi k, 2\pi (k+1)]$ interval we have $\alpha_k=\sum_{j=0}^{k-1} \beta_j$ D1 colour branes and $\mu_k=\sum_{j=0}^{k-1} \nu_j$ D5 branes. We also have $(\beta_{k-1}-\beta_{k})$ D3 and $(\nu_{k-1}-\nu_{k})$ D7 sources (flavour branes).

We close here our analysis of the new AdS$_2$ geometries. Below, we present a proposal for the  dual super conformal quantum mechanics. Matchings between holographic and field theoretical calculations, together with some holographic predictions for these quantum mechanical systems at strong coupling, are discussed in the next section.

\section{Field Theory and Holography}\label{FTH}

%
In this section we discuss the ${\cal N}=4$ super-conformal quantum mechanical theories proposed as duals to our backgrounds in eqs.(\ref{t dualised background NS})-(\ref{t dualised background RR}). 
 As anticipated in Section \ref{2dSCFT}, we provide a UV ${\cal N}=4$ quantum mechanics, that conjecturally flows to a super conformal quantum mechanics dual to our  AdS$_2$ backgrounds.

The bottomline is that the quantum mechanical quiver is the dimensional reduction of the two dimensional QFTs presented in Section \ref{2dSCFT}. Let us discuss two approaches into the quantum mechanical theory.

One approach is based on the works   \cite{Balasubramanian:2009bg,Azeyanagi:2007bj,Strominger:1998yg}. In these papers it is suggested that the transition from AdS$_3$ to AdS$_2$ should be thought  of in  CFT$_2\to$ CFT$_1$ language as a discrete light-cone  quantisation of the two dimensional CFT. This is to be taken in a limit such that, of the original SL$(2, \mathbb{R})\times$ SL$(2, \mathbb{R})$ symmetry of the seed CFT$_2$,
only one of the sectors is kept. The other sector needs  infinite energy to be excited.
Writing the boundary metric of AdS$_3$ as a cylinder, $\mathrm{d}s^2= -\mathrm{d}t^2+ \mathrm{d}\phi^2$, and changing coordinates to $u=t-\phi$ and $v=t+\phi$, we have $\mathrm{d}s^2=-\mathrm{d}u \mathrm{d}v$. In these variables the identification of coordinates
$[t,\phi]\to [t, \phi+2\pi R]$ demands $ [u,v]\to [u-2\pi R, v+2\pi R]$.
The scaling $u\to e^{\gamma}u$, $v\to e^{-\gamma}v$ keeps the metric invariant. In the limit  $\gamma\to\infty$, keeping $R e^{\gamma}=\tilde{R}$ fixed, the CFT$_2$ then lives on a space consisting on time and a null-circle. The energies scale in such a way that the left sector decouples and the right sector has $E_n=\frac{n}{\tilde{R}}$ (see  \cite{Balasubramanian:2009bg,Azeyanagi:2007bj} for the details).
The T-dualisation along the $\tilde{\psi}$-direction performed in Section \ref{A-type} is equivalent to starting with a given  ${\cal N}=(0,4)$ SCFT$_2$ as those described in Section \ref{2dSCFT} and DLCQ it, keeping the ${\cal N}=4$ SUSY right sector. Similar ideas have been discussed recently in \cite{Aniceto:2020saj}. In purely field-theoretical terms, we start with the Lagrangian alluded to  in Section \ref{2dSCFT}  (written in Appendix \ref{appendix1}) and dimensionally reduce it to a matrix model where we keep only the time dependence and the zero modes in the $\tilde{\psi}$-direction.

A second interesting way to think about our quantum mechanical theory is inspired by  the works \cite{Assel:2018rcw,Assel:2019iae}. In these references the same brane set-up depicted in Table \ref{Table brane web set type IIA} was proposed in order to describe half-BPS Wilson and 't Hooft loops in 5d gauge theories with 8 supercharges. These defects were described by quiver quantum mechanics with the same field content that we described in Section \ref{2dSCFT}, after dimensional reduction. Our quiver quantum mechanics exhibit however  additional constraints, that are inherited from the anomaly cancelation conditions of the seed 2d CFT. We will see in \cite{Lozano:2020sae},\cite{Lozano:2021rmk} that more general quivers such as the ones constructed in \cite{Assel:2018rcw,Assel:2019iae} are dual in the IR to AdS$_2$ solutions not related to AdS$_3$ upon T-duality.

 %
 
In summary, our proposal is that the dynamics of the  UV quantum mechanical systems of interest, is described by the dimensional reduction along the space-direction of the  ${\cal N}=(0,4)$ SCFT$_2$ discussed in Section \ref{2dSCFT}. 
To be concrete: consider the Type IIB backgrounds described in eqs.(\ref{t dualised background NS})-(\ref{t dualised background RR}) with the functions $\hat{h}_4, h_8$ given in eqs. (\ref{profileh4sp})-(\ref{profileh8sp}). These solutions are dual to an ${\cal N}=4$ superconformal quantum mechanics that arises in the IR of a generic quiver quantum mechanics, with the matter content depicted in Figure \ref{figurageneral}. The dynamics is inherited from the two-dimensional ${\cal N}=(0,4)$ Lagrangian by dimensional reduction. We can read the ranks of colour and flavour groups from the Page charges computed in eqs.(\ref{cargaspageA}). In the $k^{th}$ entry, corresponding with the $[2\pi k, 2\pi (k+1)]$ interval of the geometry, we have U$(\alpha_k)$ and U$(\mu_k)$ colour groups---with $\alpha_k=\sum_{j=0}^{k-1}\beta_j,\; \mu_k=\sum_{j=0}^{k-1}\nu_j$. These are coupled via bifundamental hypermultiplets  and Fermi multiplets with the  adjacent nodes. The connections with the $k^{th}$ flavour groups of ranks SU$(\nu_{k-1}-\nu_{k})$ and SU$(\beta_{k-1}-\beta_{k})$ is mediated by Fermi fields and by bifundamental hypermultiplets.

The authors of  \cite{Assel:2019iae} impose that the numbers of D3 and D7 (sources/flavour) branes must equal the difference of two integers. In our formalism this is automatic. The integers are identified with the ranks of the colour groups (be it D1 or D5) on each side of the interval. We have that the number of D3 sources is $(\beta_{k-1}-\beta_k)$ and analogously  $(\nu_{k-1}-\nu_{k})$ for the number of D7 flavours. These numbers are positive as guaranteed by the convex character of our polygonal functions  $\hat{h}_4, h_8$. The quiver is identical to and inherited from that of the  two-dimensional ``mother'' theory--see  Figure \ref{figurageneral}. The superfields involved in writing the Lagrangian are also inherited, as explained in Appendix \ref{appendix1}.

In what follows, we perform some holographic calculations that inform us about the strong dynamics of these conformal quantum mechanical quivers.
\subsection{The holographic central charge}\label{hccsection}
The definition of central charge in conformal quantum mechanics is subtle.
 In a one-dimensional theory, we have only one component of $T_{\mu\nu}$. If the theory is conformal, the trace of this quantity must vanish, and this implies that $T_{tt}=0$.  One way to think about central charge  is to consider a conformal  quantum mechanics which has many ground states (but no excitations). One may associate this quantity with the ``central extension'' of the Virasoro algebra that appears in the global charges of the two-dimensional dual gravity, as discussed in \cite{Hartman:2008dq,Alishahiha:2008tv,Cadoni:1999ja}. We can also associate this ``central charge''  with the partition function of the quantum mechanics when formulated on a circle, as discussed for example in \cite{vanBeest:2020vlv}. 

Though we refer to it as ``holographic central charge'' the quantity that we present below should be interpreted as a ``number of vacuum  states'' in the associated SCQM. To define it, we shall use the logic discussed in
\cite{Klebanov:2007ws,Macpherson:2014eza}.

We follow the prescription in   \cite{Macpherson:2014eza}. Being the field theory  zero-dimensional, some of the steps in the calculation need some care. The relevant quantity in this case is the volume of the internal space (the part not belonging to AdS$_2$). Analogously, we are  computing Newton's constant in an effective two-dimensional gravity theory,
\begin{equation}
\frac{1}{G_{N,2}}=\frac{V_{int}}{G_{N,10}}.\label{4.0}
\end{equation}
Following the formalism of \cite{Macpherson:2014eza} for the backgrounds described in eqs.(\ref{t dualised background NS})-(\ref{fluxes2}), we find
\begin{equation} 
V_{int}=\int \mathrm{d}^8 x \sqrt{e^{-4\Phi} \det g_{8,ind}} =\left( \frac{    \text{Vol}_{\text{CY$_2$} }    \text{Vol}_{\text{S$^2$} }  \text{Vol}_\psi    }{4} \right)\int_0^{2\pi(P+1)} \hat{h}_4 h_8 \text{d}\rho.\label{chola}
\end{equation}
The comparison with eq.(\ref{chol}) indicates that, under  suitable rescaling,  this quantity is  related to the central charge of the seed  2d SCFT.

We define the ``holographic central charge'' of the conformal quantum mechanics to be
\begin{eqnarray}
c_{hol,1d}=\frac{3}{4\pi G_{2}}= \frac{3 V_{int}}{4\pi G_{N}}.\label{definitioncentral}
\end{eqnarray}
Computing explicitly with eq.(\ref{chola}) and using that (in the units $g_s=\alpha'=1$) $G_{N}= 8 \pi^6$, we find
\begin{equation}
c_{hol,1d}=\frac{3}{\pi}\int_0^{2\pi(P+1)} h_4h_8 \text{d}\rho, \label{cholA}
\end{equation}
 in agreement with the two-dimensional result in eq.(\ref{chol}). This is compatible with the findings of the paper \cite{Balasubramanian:2009bg}, that suggest that the chiral sector remaining when DLCQ is applied to a 2d CFT has the same central extension in the Virasoro algebra.

On purely field theoretical terms, this result tells us that the number of vacua of the  ${\cal N}=4 $ SCQM obtained by dimensional reduction  of the two-dimensional ``mother theory'', responds to the expression obtained in
\cite{Putrov:2015jpa}, namely 
\begin{equation}
c_{qm}=6(n_{hyp}- n_{vec}).
\label{central1d}
\end{equation}
 The numbers of ${\cal N}=4$ hyper and vector multiplets in the one dimensional theory are inherited from those in the two dimensional ``mother'' theory. The agreement between $c_{qm}$ in eq.(\ref{central1d}) and $c_{CFT}$ in eq.(\ref{centralcft}) is the field theoretical translation of the equality of the holographic central charges in two dimensions, eq.(\ref{chol}), and in one dimension, eq.(\ref{cholA}).

\begin{figure}[h!]
    \centering
    {{\includegraphics[width=10cm]{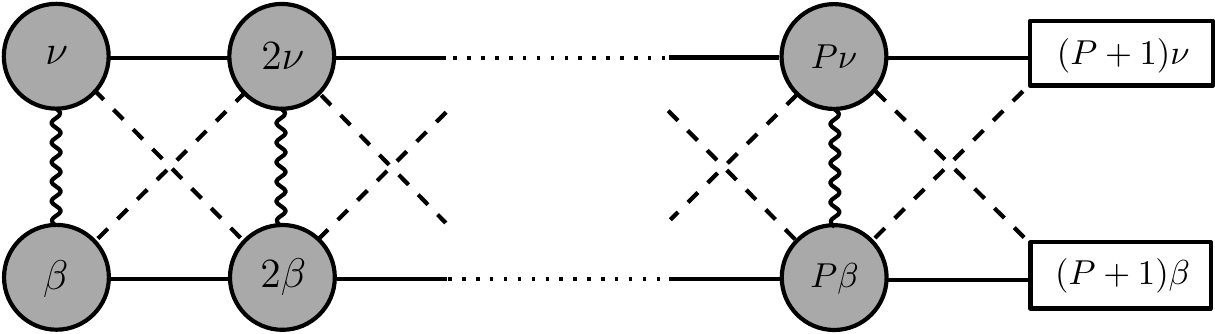} }}%

\caption{ The quantum mechanical system that conjecturally flows in the IR to the SCQM described by the backgrounds obtained from eqs.(\ref{profileh8exampleII})-(\ref{profileh4exampleII}).}
\label{Example}
\end{figure}

It is interesting to draw a comparison with  the works \cite{Denef:2002ru,Ohta:2014ria,Cordova:2014oxa}. These papers make crucial use of the dimension of the Higgs branch for a quiver quantum mechanics with gauge group $\Pi_{v} U(N_v)$ and bifundamentals joining each colour group with
the adjacent ones. This quantity is given by,
\begin{equation}
{\cal M}=\sum_{v,w} N_v N_w -\sum_v N_v^2 +1.\label{dimhiggs}
\end{equation}
We propose that the calculation in eq.(\ref{cholA}) captures the same information as eq.(\ref{dimhiggs}).  Note that our quantum mechanical theories have a field content that is more involved than the ones  considered in \cite{Denef:2002ru,Ohta:2014ria,Cordova:2014oxa}.
Let us illustrate this with an example (similar calculations can be done for other quivers). The example we choose is represented by the functions,
\begin{equation} \label{profileh8exampleII}
h_8(\rho)
                    =\left\{ \begin{array}{ccrcl}
                       \frac{\nu }{2\pi}
                       \rho & 0\leq \rho\leq 2\pi P \\
                      \frac{\nu P}{2\pi}(2\pi( P+1) -\rho), & 2\pi P\leq \rho \leq 2\pi(P+1).
                                             \end{array}
\right.
\end{equation}
  \begin{equation} \label{profileh4exampleII}
\hat{h}_4(\rho)=\Upsilon h_4(\rho)
                    =\Upsilon\left\{ \begin{array}{ccrcl}
                       \frac{\beta }{2\pi}
                       \rho & 0\leq \rho\leq 2\pi P\\
                      \frac{\beta P}{2\pi}(2\pi (P+1) -\rho), & 2\pi P\leq \rho \leq 2\pi(P+1).
                                             \end{array}
\right.
\end{equation}
According to the rules presented in Section \ref{FTH}, the ${\cal N}=4$ quantum mechanical quiver is the one depicted in Figure \ref{Example}.
We  calculate the expressions for the one dimensional central charge $c_{qm}$ in eq.(\ref{central1d})  and its holographic counterpart $c_{hol,1d}$ in eq. (\ref{cholA}). These expressions should coincide in the holographic limit with the dimension of the Higgs branch in eq.(\ref{dimhiggs}). Using the definitions in eqs.(\ref{cholA}) and (\ref{central1d}) we calculate
\begin{eqnarray}
& & n_{hyp}= \sum_{j=1}^P \Bigl(j(j+1)(\nu^2+\beta^2) + j^2 \nu \beta\Bigr),\;\;\;\; n_{vec}=\sum_{j=1}^P j^2 (\nu^2+\beta^2),\label{centralex}\\
& & c_{qm}=  3 P (P+1)(\nu^2 + \beta^2) + (2 P + 1)(P+1)P \nu \beta \sim 2\beta\nu P^3 \nonumber  \\
& & c_{hol,1d}=2\beta\nu (P^3 + P^2)\sim 2\beta\nu P^3.\nonumber
\end{eqnarray}
We see that in the holographic limit (large $P, \nu,\beta$) the results  of eqs.(\ref{cholA}) and (\ref{central1d}) coincide.
At the same time we see that the dimension of the Higgs branch moduli space in eq.(\ref{dimhiggs}) is precisely counting the number of hypers minus the number of vectors. Note  that our quiver has hypers joining the links not only  ``horizontally'' but also ``vertically'', in comparison with the quivers considered in  \cite{Denef:2002ru,Ohta:2014ria,Cordova:2014oxa}.
 
Following  \cite{Lozano:2019zvg}-\cite{Lozano:2019ywa}, the reader can produce a variety of test-examples showing the coincidence of the calculations of eqs.(\ref{cholA}), (\ref{central1d}) and (\ref{dimhiggs}) in the holographic limit (it should be interesting  to explore sub-leading corrections!). We shall come back to the holographic central charge and relate it to an extremisation principle in Section \ref{HCCMP}.

Let us now discuss  predictions for the strong coupling dynamics of our SCQMs.
\subsection{Chern-Simons terms}
Let us discuss the possible ``dynamical''  term for the gauge multiplet. In $(0+1)$ dimensions this is a Chern-Simons (CS) term. Let us motivate their presence with a small detour on anomalies.

 The authors of \cite{Elitzur:1985xj} present a detailed study on the conflict between gauge symmetry and global symmetry (charge conjugation in this case). They study the action of $l$-fermions on a time circle of size $T$, in the presence of a $U(1)$ gauge field $A_t(t)$. The system has Lagrangian, gauge and charge conjugation transformations given by,
\begin{eqnarray}
& & L= \bar{\psi}(i \partial_t + A_t)\psi.\nonumber\\
& & \psi\to e^{i \Lambda}\psi,\;\; \bar{\psi}\to \bar{\psi}e^{-i \Lambda},\;\;\; A_t\to A_t+\partial_t \Lambda(t), \;\;\;\; A_t\to -A_t.\label{transformations}
\end{eqnarray}
For  configurations that are periodic in the circle,  the partition function (for $l$ fermions with the above Lagrangian) is,
\begin{eqnarray}
& & Z= \int {D}\psi D\bar{\psi} e^{-i \int_0^T \mathrm{d} t L}= \det(i \partial_t + A_t)^l= (1+ e^{i a_0 T})^l,\label{partition}\\
& & a_0 T=\int_0^T A_t(t) \mathrm{d} t.\nonumber
\end{eqnarray}
This is invariant under both large and small gauge transformations, but not under charge conjugation. A way to recover the charge conjugation invariance is through the introduction of a counterterm
\begin{equation}
L_{ct}= e^{-i k \int_0^T A_t \mathrm{d} t}= e^{-i k a_0 T} . \label{counterterm}
\end{equation}
This  is a CS-term. In itself, it is gauge invariant but not charge conjugation invariant. If $2 k =l$,  its presence cancels the lack of invariance under charge conjugation in eq.(\ref{partition}). We can regularise the partition function of  an even number of fermions, such that gauge invariance and charge conjugation are both preserved.
 If the number of  fermions is odd, we just loose the charge conjugation invariance.
 
In  summary, for the case of $(0+1)$-dimensions the Chern-Simons term  is of the form
\begin{equation}
S_{CS}=\kappa_{CS}\int \mathrm{d}t A_t.\nonumber
\end{equation}
The coefficient $\kappa_{CS}$ must be quantised. As above, consider the theory on a circle of length $T$. Performing a large gauge transformation, $A_t\to A_t+ \partial_t \Lambda$ with parameter $\Lambda= \frac{2\pi n }{T}t$, we find that the Chern-Simons action changes,
\begin{equation}
S_{CS}\to S_{CS} +\kappa_{CS} 2\pi n.\nonumber
\end{equation}
Imposing that $e^{i S_{CS}}$ is single-valued under large gauge transformations, we find that $ e^{i 2\pi n \kappa_{CS}}=1$, which quantises the Chern-Simons coefficient.
\subsubsection{Holographic calculation of the Chern-Simons coefficients}
Let us holographically compute the Chern Simons coefficients
for each gauge group in the quantum mechanical quiver derived by dimensional reduction of that in Figure \ref{figurageneral}. The presence of the CS term is of non-perturbative origin. We calculate it using the Type IIB AdS$_2$  description of the system. To do so we use a D1 brane probe extended in $[t,\rho]$, with a gauge field (of curvature $F_{t\rho}$) excited on it. The Wess-Zumino term for the D1 brane probe reads,
\begin{eqnarray}
S_{WZ}\!=\!T_{\text{D1}}\!\!\int C_p\wedge e^{2\pi F_2}=\! T_{\text{D1}}\! \left(\int C_2\! +\! 2\pi \int C_0 \!F_{t\rho} \mathrm{d} t \text{d}\rho\right)= -2\pi T_{\text{D1}}\!\! \int \!\mathrm{d}t\! \int\! \text{d}\rho A_t \partial_{\rho}C_0.\label{swzcs}
\end{eqnarray}
In the last equality we have used that the RR field $C_2$ has no pull-back on this probe. Moreover, we have  set the gauge $A_{\rho}=0$  and imposed that the gauge field $A_t$ takes the same values at the extrema of the interval. Keeping in mind that the axion field $C_0$ is only well-defined in regions where $h_8'$ is a constant\footnote{Due to the presence of sources, see eqs.(\ref{fluxes2})-(\ref{caxa}), the axion field is not globally defined.}, where it reads
$C_0= h_8' \psi$, we find,
\begin{equation}
S_{WZ}=-2\pi T_{\text{D1}}\int \mathrm{d}t \text{d}\rho A_t(t,\rho) \psi h_8''= -2\pi T_{\text{D1}}\psi (\nu_{k-1}-\nu_{k}) \int \mathrm{d}t A_t(t, 2\pi k)=\kappa_{CS,I} \int A_t \mathrm{d}t.
\end{equation}
Using eq.(\ref{swzcs}) we have that the Chern-Simons coefficient  in the interval $[2\pi k, 2\pi(k+1)]$ is then given by,
\begin{eqnarray}
& & \kappa_{CS, I}[k,k+1 ]=\psi\frac{  (\nu_{k-1}-\nu_{k})}{2\pi}.\label{kappacs}
\end{eqnarray}
Therefore, in order to keep the CS coefficient well quantised, we can allow discrete  changes of the  coordinate $\psi$,
 \begin{equation}
 \psi\to\psi + \left( \frac{2\pi l }{\nu_{k-1}-\nu_{k}}\right),~~ \text{with} ~~ l=1, \dots , (\nu_{k-1}-\nu_{k}) .
 \end{equation} 
 These changes indicate that not all positions in $\psi$ are allowed for the D1 probes. 
 In other words, the $U(1)_\psi$ isometry of the background is broken  to $\mathbb{Z}_{\nu_{k-1}-\nu_{k}}$.
On the other hand, the presence of the source D7 branes implies a change in the Chern-Simons coefficient, as the slopes of the function $h_8$ change. 

A very similar calculation for a D5 brane that extends on $[t,\rho, \text{CY}_2]$ gives a Chern Simons coefficient for the gauge groups in the lower row that is
\begin{equation}
\kappa_{CS, II}[k, k+1]=\psi\frac{(\beta_{k-1}-\beta_{k})}{2\pi}.\label{csiix}
\end{equation}
We find that the $U(1)_\psi$ is broken to $\mathbb{Z}_{\beta_{k-1}-\beta_{k}}$ and $\mathbb{Z}_{\nu_{k-1}-\nu_{k}}$, by the Chern-Simons terms in the lower and upper rows respectively.
If they have no common subgroups  the $U(1)_\psi$ is completely broken.
Notice also that the sum of all the Chern-Simons coefficients gives $\sum_k \kappa_{CS}[k,k+1]= N_F \psi$, where $N_F$ is the sum of the total number of D7 brane sources in the upper row and the total number of D3 branes sources  in the lower row. 

These are non-trivial predictions for the strongly coupled dynamics of our ${\cal N}=4$ SCQM. In Appendix \ref{othercalculations} we discuss additional ones.
We now  go back to discussing the holographic central charge from two different perspectives.

\section{Holographic central charge, electric-magnetic charges and a minimisation principle}\label{HCCMP}
In this section we present different ways of understanding the holographic central charge given by  eq.(\ref{cholA}).
We give a two-fold presentation. In  Section \ref{pagexxxx}, that is more physically inspired, we show that the expression in eq.(\ref{cholA}) is related to
a product of electric and magnetic charges associated with our backgrounds. In Section \ref{HCCMP2} we present a more geometrical approach, finding that the expression (\ref{cholA}) can be obtained via an extremisation principle.
\subsection{The relation between central charge and Page fluxes}\label{pagexxxx}
We  study the link between the holographic central charge of the SCQM in eqs.(\ref{chola})-(\ref{cholA}) with the integral of electric and magnetic fluxes in the ten dimensional space. We  see this explicitly by working with the Page fluxes in eqs.(\ref{fluxes2}).

 This calculation is the string theoretic realisation of an argument presented in \cite{Hartman:2008dq} for two dimensional AdS$_2$ gravity. In this reference it was proposed  that the central charge of the SCQM should be related to the (square of the) electric field in an effective AdS$_2$ gravity theory coupled to a gauge field. 

Consider a Dp brane, to which we can associate electric  $\hat{F}_{p+2}$ and magnetic $\hat{F}_{8-p}$ Page field strengths. We define the ``density of electric and magnetic charges'',  $\rho^e_{\text{Dp}}$ and $\rho^m_{\text{Dp}}$, as the forms
\begin{eqnarray}
& & \rho_{\text{Dp}}^{e}=\frac{1}{(2\pi)^p} \hat{F}_{p+2} ,\;\;\;\;\; \rho_{\text{Dp}}^{m}=\frac{1}{(2\pi)^{7-p}} \hat{F}_{8-p}.\label{mc}
\end{eqnarray}
The electric charge, obtained by integration of the charge density form, will turn out to be infinite, as it involves the integration of the volume form of the non-compact AdS$_2$ spacetime. We will work with these definitions, having in mind that a regularisation will be necessary after the integrations are performed, see for example \cite{Aguilera-Damia:2017znn}. 

Consider the product of electric and magnetic charge densities in eq.(\ref{mc}), and its integration over all space for the D-branes present in our backgrounds. We show that (after regularisation) this product is proportional to the holographic central charge given by eq.(\ref{cholA})\footnote{In order to show this we use that only one of the components of $\hat{F}_5\wedge \hat{F}_5$ needs to be taken into account, due to its self-duality, and that some sign flips are necessary in order to work with the absolute values of the charges and avoid unwanted cancellations.}. We 
calculate the integral of electric and magnetic densities in eq.(\ref{mc}) using the Page fluxes derived in eq.(\ref{fluxes2}), and the ordered basis $[t,r,\text{S}^2, \textrm{CY$_2$},\rho,\psi]$. The calculation leads to
\begin{eqnarray}
& & \int \sum_{k=0}^3 (-1)^k \rho_{\text{D}(2k+1)}^e \rho_{\text{D}(2k+1)}^m=\label{ppaa}\\
& & =\int \text{d}\rho \left(\frac{ \hat{h}_4h_8}{2} +\frac{1}{16}\partial_\rho\left[2 u u' - u^2\left(\frac{(\hat{h}_4 h_8)'}{\hat{h}_4h_8} \right) \right] \right)\text{Vol}_{\text{AdS$_2$}} \left( \frac{\text{Vol}_{\text{S$^2$}} }{4\pi^2}\right) \left(\frac{\text{Vol}_{\text{CY$_2$}} }{16\pi^4} \right) \left( \frac{\text{Vol}_\psi}{2\pi}\right).\nonumber 
\end{eqnarray}
 Up to a boundary term, this is  proportional to eq.(\ref{cholA}), the expression for the holographic central charge of our AdS$_2$ backgrounds.

Hence, we learn that the holographic central charge in eq.(\ref{cholA}), measuring the number of vacua of the associated SCQM, is proportional to the (regularised) product of  electric and magnetic charge densities. We see this relation as a generalisation of the proposal in  \cite{Hartman:2008dq}, showing that the central charge in the algebra of symmetry generators of AdS$_2$ with an electric field is proportional to the square of the electric field. In our case, for a fully string theoretic set-up, we have objects with electric and magnetic charges and both enter the calculation.

This  links the holographic central charge, usually calculated from the dilaton and the metric of the internal space, as shown by equations (\ref{4.0}) and (\ref{chola}), with an expression purely in terms of the RR-sector. It would be nice to see a similar logic at work in higher dimensional AdS-solutions.

\subsection{An Extremisation Principle}\label{HCCMP2}
In this section we present  a minimisation principle in supergravity that will lead to the expression for the holographic central charge in eq.(\ref{cholA}).
Our presentation falls in line with the ideas that extremisation problems in quantum field theory are realised in supergravity through the extremisation of certain geometrical quantities.
Various examples exist of this mirroring of extremal principles. The most relevant to us are the ones studied in  \cite{Couzens:2018wnk,Hosseini:2019ddy,vanBeest:2020vlv}. In these papers a geometrical quantity is defined in supergravity that coincides upon extremisation with the holographic central charge of the systems under study. In some cases this defines the central charge of the dual field theory. We point out some extensions and differences with the approach of  \cite{Couzens:2018wnk,Hosseini:2019ddy,vanBeest:2020vlv}.

Let us follow the idea of  \cite{Couzens:2018wnk}. 
These authors  consider a particular family of backgrounds (in eleven dimensional supergravity)
of the form AdS$_2$ $\times$ Y$_9$, containing an electric flux $F_4$ and preserving ${\cal N}=(0,2)$ SUSY. Aside from the AdS$_2$ factor, these backgrounds are quite different from the ones we discuss here (or their lift to M-theory,  in the case in which $h_8$  is constant \cite{Lozano:2020bxo}). Nevertheless, the lesson from 
 \cite{Couzens:2018wnk} is that the central charge can be written in terms of an extremised functional. This functional is defined as an integral of various forms in the geometry Y$_9$, and it is such that, once extremised, equals a weighted volume of the internal space. Importantly, the manifold Y$_9$ in  \cite{Couzens:2018wnk} has no boundary. In our case, we have a boundary and we allow for the presence of sources.
 
In order to implement these ideas we define certain differential forms on the $X_8$ internal space tranverse to our AdS$_2$ solutions, $X_8=[S^2, \text{CY}_2,\rho,\psi]$. We construct these forms {\it restricting}
the Page forms in eq.(\ref{fluxes2}) to the manifold $X_8$.  For example, from $\hat{F}_1$ we generate the one-form 
\begin{equation}
\hat{F}_1\longrightarrow J_1= h_8' \text{d}\psi.\label{J1xx}
\end{equation}
From the Page form $\hat{F}_3$ in eq.(\ref{fluxes2}) we generate a second one-form, plus a three-form,
\begin{eqnarray}
& & \hat{F}_3\longrightarrow  {\cal F}_1= \left(\frac{h_8}{2} +\frac{u'^2 \hat{h}_4 - u u' \hat{h}_4}{8 \hat{h}_4^2}\right) \text{d}\rho,\;
 J_3= -\frac{1}{2}(h_8- h_8'(\rho-2\pi k)) \text{vol}_{\text{S$^2$}}\wedge \text{d}\psi.\label{F1J3xx}
\end{eqnarray}
The other forms generated from the Page fluxes are,
\begin{eqnarray}
& &  {\cal F}_3=  \frac{1}{16} \bigg(  \frac{\left(u -( \rho-2\pi k)  u' \right) ( u  \hat{h}_4' - \hat{h}_4  u'  )}{ \hat{h}_4^2} + 4( \rho-2\pi k)  h_8     \bigg) \text{vol}_{\text{S$^2$}}\wedge \text{d}\rho,\nonumber\\
& & J_5= - \hat{h}_4'\text{vol}_{\textrm{CY$_2$}}  \wedge \text{d}\psi,\;\;\;\; {\cal F}_5= -\left(  \frac{4 \hat{h}_4 h_8^2 - u u' h_8' + h_8 (u')^2}{8 h_8^2} \right) \text{vol}_{\textrm{CY$_2$}}  \wedge \text{d}\rho,\nonumber\\
& &  J_7= \frac{1}{2}(\hat{h}_4- \hat{h}_4'(\rho-2\pi k)) \text{vol}_{\textrm{CY$_2$}}  \wedge  \text{vol}_{\text{S$^2$}}\wedge \text{d}\psi,\label{formasxx}\\
& & {\cal F}_7=  - \left(\frac{ 4 (\rho-2\pi k) \hat{h}_4 h_8^2 + u^2 h_8' - h_8 u u' -(\rho-2\pi k) u u' h_8' +(\rho-2\pi k) h_8 u'^2}{16 h_8^2}\right) \text{vol}_{\textrm{CY$_2$}}  \wedge  \text{vol}_{\text{S$^2$}}\wedge \text{d}\rho.\nonumber
\end{eqnarray}
With the forms in eqs.(\ref{J1xx})-(\ref{formasxx}), we define the functional,
\begin{eqnarray}
& & {\cal C}= i\int_{X_8} (J_3+ i {\cal F}_3)  \wedge (J_5+ i {\cal F}_5) - (J_1 + i {\cal F}_1)\wedge (J_7+ i {\cal F}_7)= \label{functionalc}\\
& & = \frac{1}{16}\int_{X_8}  \left({8\hat{h}_4 h_8}{} + u^2 \left(\frac{\hat{h}_4'^2}{\hat{h}_4^2}+\frac{h_8'^2}{h_8^2} \right)- 2 u u' \left(\frac{\hat{h}_4'}{\hat{h}_4} +\frac{h_8'}{h_8} \right) + 2 u'^2  \right)\text{vol}_{\textrm{CY$_2$}}  \wedge  \text{vol}_{\text{S$^2$}}\wedge \text{d}\rho\wedge \text{d}\psi.\nonumber
\end{eqnarray}
Let us remind that the functional ${\cal C}$ is defined in terms of the {\it restriction} on $X_8$ of the Page fluxes.
This can be minimised by imposing the Euler-Lagrange equation  for $u(\rho)$ from the ``Lagrangian''
in eq.(\ref{functionalc}). This equation reads
\begin{eqnarray}
& & 2 u''=u\left(\frac{\hat{h}_4''}{\hat{h}_4}+\frac{h_8''}{h_8}\right).
\end{eqnarray}
Imposing the Bianchi identities
\begin{equation}
h_8''=0,\;\;\;\;\;\hat{h}_4''=0,\label{EL}
\end{equation}
this leads us to the BPS equation of our class of solutions (\ref{t dualised background NS})-(\ref{t dualised background RR})\footnote{Below, we analyse the situation in which sources are present. This implies the presence of delta-function sources as in eq.(\ref{nana}). },
\begin{equation}
u''=0.\label{equxx}
\end{equation} 
Note that the fluxes are quantised, due to the type of solutions we consider for $\hat{h}_4, h_8$--see eqs.(\ref{cargaspageA}). It is interesting that here we impose the Bianchi identities in eq.(\ref{EL}) leading to the BPS eq.(\ref{equxx}). This is different from the procedure followed in previous bibliography.

On the solutions to eqs. (\ref{EL}), (\ref{equxx}), which we refer to as ``on-shell',  the functional is extremised to be,
\begin{eqnarray}
& & {\cal C}|_{on-shell}=\left( \frac{    \text{Vol}_{\text{CY$_2$} }    \text{Vol}_{\text{S$^2$} }  \text{Vol}_\psi    }{2} \right)\int_0^{2\pi(P+1)} \left(\hat{h}_4 h_8  + \partial_\rho {\cal M}\right)\text{d}\rho.\label{comparewith}\\
& & \text{with} \quad {\cal M}=\frac{1}{8}\left( 2 u u'- u^2\Bigl(\frac{\hat{h}_4'}{\hat{h}_4}+\frac{h_8'}{h_8}\Bigr)\right).\nonumber
\end{eqnarray}
We now compare equations (\ref{cholA}) and (\ref{comparewith}). Up to a boundary term (present in $X_8$ but not in the boundary-less $Y_9$ of  \cite{Couzens:2018wnk}),  the inverse Newton's constant in two dimensions in eq.(\ref{4.0}), the internal volume $V_{int}$ in eq.(\ref{chola}) and the ${\cal C}|_{on-shell}$ in eq.(\ref{comparewith}),
  all converge into the same calculation. Our proposed duality implies that these geometrical quantities count  the number of vacua of the dual SCQM, according to eqs.(\ref{cholA})-(\ref{central1d}).

Let us come back to eq.(\ref{functionalc}), and analyse it for the case in which we have sources. After integration by parts we write it  as,
\begin{eqnarray}
& &{\cal C}= \frac{1}{16}\int_{X_8}  \left({8\hat{h}_4 h_8}{} + {\cal N} +   \partial_\rho {\cal M}\right)\text{d}\rho,\nonumber \qquad \text{with} \\
& & {\cal N}= u^2\left(\frac{\hat{h}_4''}{\hat{h}_4}+\frac{h_8''}{h_8} \right),\;\;\;\; {\cal M}= 2 u u'- u^2\left(\frac{\hat{h}_4'}{\hat{h}_4}+\frac{h_8'}{h_8}\right).\nonumber
\end{eqnarray}
Using eq.(\ref{nana}), the term ${\cal N}$ can be seen
to give a finite result proportional to the quotient of the number of flavours by the number of colours in each node. Let us understand the boundary term in more detail. For the solutions in eqs.(\ref{profileh4sp})-(\ref{profileh8sp}) the boundary term  is divergent.
We can associate this divergence with the presence of sources. Consider for example a solution for ${\hat{h}_4}, h_8$ of the type in eqs.(\ref{profileh4sp})-(\ref{profileh8sp}) with $u=u_0$ (a constant). In that case evaluating the boundary term we find
\begin{equation}
\int_0^{2\pi(P+1)}\partial_\rho {\cal M}=\lim_{\epsilon\to 0} \frac{u_0^2}{\epsilon}\left(\nu_0+ \beta_0 +\alpha_P +\mu_P \right)=\lim_{\epsilon\to 0} \frac{u_0^2}{\epsilon}\left(N_{\text{D3}}^{total} + N_{\text{D7}}^{total}\right).
\end{equation}
We have used that $\hat{h}_4(\rho=0)=h_8(\rho=0)=\epsilon$ and the same for the  corresponding values at $\rho=2\pi(P+1)$. Then we take $\epsilon\to 0$, finding a divergent result in terms
of the total number of sources present in the background.
Notice that in the limit of long quivers ($P$ large) with large ranks for colour group nodes $U(\alpha_k)$ and $U(\mu_k)$ and sparse flavour groups, both the boundary term ${\cal M}$ and the bulk term ${\cal N}$, conveniently renormalised, are subleading in these numbers ($P, \alpha_k,\mu_k$) with respect to the first term, proportional to the holographic central charge in eq.(\ref{cholA}). For this we need to define the functional ${\cal C}$ in eq.(\ref{functionalc}) with a suitable counter-term that removes the divergence when $\epsilon\to 0$.

It should be interesting to attempt the calculation presented here in different systems, like those in \cite{Lozano:2019jza} or in higher dimensional AdS-spaces, to check if similar extremisation principles are at work. In particular, it would be interesting to understand the geometrical meaning of the forms  in eqs.(\ref{J1xx})-(\ref{formasxx}).

In summary, the presentation above shows  that the holographic central charge, originally defined purely in terms of the NS-NS sector---see eqs.(\ref{chola})-(\ref{cholA}), is  also encoded in the forms of eq.(\ref{formasxx}) and the functional (\ref{functionalc}). The contents of this section link the holographic central charge with the product of electric and magnetic brane charges and with an extremisation principle.
These geometrical quantities are capturing the number of vacua of the ${\cal N}=4$ SCQM.


\section{Summary and Conclusions}\label{closing}
Given that this is a long and dense paper, the reader may find useful to start with a summary. We describe the main new ideas and calculations presented, pointing to the sections and  equations that best describe them.

We start in Section \ref{geometria} with a summary of the seed-backgrounds in massive IIA, dual to two-dimensional ${\cal N}=(0,4)$ SCFTs. The new material is written in Section \ref{2dSCFT}. There we discuss in detail the field content of the two dimensional field theories. Also, in Section \ref{secsupxx} we presented the superpotential for these two-dimensional field theories. Details and generalisations are given in Appendices \ref{appendixgeneral},\ref{appendix1}.

In Section \ref{A-type}, we have constructed new AdS$_2$ solutions to Type IIB supergravity with $\mathcal{N}=4$ supersymmetry. 
This infinite family of solutions is precisely written in eqs.(\ref{t dualised background NS})-(\ref{nana}).  The Page charges are calculated and the Hanany Witten set-up summarised in Table \ref{Table brane web set type IIA}.

In Section \ref{FTH}, we propose explicit quiver quantum mechanics that should flow in the IR to the SCQM dual to the backgrounds in Section \ref{A-type}.
Some aspects of the dynamics of the SCQM have been calculated using the dual description. For example the number of vacua, that we equated with the holographic central charge of the SCQM. This quantity is derived in eqs.(\ref{4.0})-(\ref{cholA}). Our expressions are tested with an example in eqs.(\ref{profileh8exampleII})-(\ref{centralex}), showing the precise match between a field theory and holography calculations.
The holographic central charge is in turn identified with the partition function of the quantum mechanics when formulated on a circle  \cite{vanBeest:2020vlv}. We have seen that this quantity is related to the 
``seed'' two-dimensional ${\cal N}=(0,4)$ SCFT right-moving central charge. This is an explicit manifestation of the DLCQ upon which both CFTs are related.
In this section we also presented  predictions for the conformal quantum mechanics. For example, we calculated the 
 Chern-Simons coefficients in eqs.(\ref{swzcs})-(\ref{csiix}). This lead to a prediction for  the number of vacua and the anomalous breaking of the symmetry $U(1)_\psi$. Wilson loops, baryon vertices and gauge couplings  have been studied in Appendix  \ref{othercalculations}.

In Section \ref{HCCMP}, we  link the holographic central charge (a quantity originally defined in terms of the NS-NS sector of the solutions) to the RR sector of our AdS$_2$ backgrounds. In particular, we have shown that it is related to the integral of the product of the electric and magnetic charge densities of the D-branes present in the system--see eq.(\ref{ppaa}). This generalises the proposal in \cite{Hartman:2008dq}, where the central charge in the algebra of symmetry generators of AdS$_2$ is related to the square of the electric field. In our controlled string theory set-up, all electric and magnetic charges of the D-branes present in the solution enter the calculation. 
In this same section, we have presented an extremisation principle following the general ideas about geometric extremisation in \cite{Couzens:2018wnk,Hosseini:2019ddy,vanBeest:2020vlv},  from which we have derived the holographic central charge. Our extremising functional is constructed in terms of the electric and magnetic RR fluxes associated to the solutions, see eqs.(\ref{J1xx})-(\ref{functionalc}). Our results extend those in \cite{Couzens:2018wnk,Hosseini:2019ddy,vanBeest:2020vlv}, by the presence of sources and boundaries.
\\
Let us end with some proposed research for the future. It would be interesting to see if a similar relation between the holographic central charge and products of Ramond fluxes, holds for other classes of solutions, especially higher dimensional ones. That would allow for a physical principle underlying the construction of the purely geometric extremising functional. Moreover, it would be interesting to find a field theory interpretation for the extremisation construction found for our AdS$_2$ solutions. Being the R-symmetry non-Abelian it is not clear why an extremisation should be necessary in order to identify the right R-symmetry from which the central charge should be constructed. This issue clearly deserves a more careful investigation. 

It would be interesting to relate, in the holographic regime, the calculations of an index (at leading order) with our holographic central charge. More generally, it would be interesting to apply exact calculation techniques to our quiver  quantum mechanics in order to understand the properties of the SCQM in the infrared. Related to this is the possibility of learning about supergravity using exact results, along the lines of \cite{GonzalezLezcano:2020yeb}.

It would be interesting to find a defect interpretation for our AdS$_2$ solutions, possibly along the lines in \cite{Faedo:2020nol}. In this reference the AdS$_3$ ``seed'' solutions from which our AdS$_2$ solutions have been constructed were interpreted as surface defects within the 5d Sp(N) gauge theory dual to the AdS$_6$ Brandhuber-Oz background \cite{Brandhuber:1999np}. It is likely that, upon T-duality, our solutions would find a similar interpretation, this time in terms of line defects, within the T-dual of the Brandhuber-Oz background \cite{Lozano:2012au,Lozano:2018pcp}. It would be interesting to find a flow interpolating these solutions with this AdS$_6$ background. These flows may be found in six dimensional supergravity, like those in \cite{Couzens:2019wls,Faedo:2020nol,Dibitetto:2017klx,Dibitetto:2018iar,Dibitetto:2018gtk} .

It should be possible to try to find compactifications of Type IIB supergravity to AdS$_2$ times an eight manifold, of the form $M_8=\text{S}^2\times \textrm{CY$_2$}\times \text{S}^1_{\psi}\times \text{I}_\rho$. Having these gauged supergravities may allow to study flows away from AdS$_2$, along the lines of those in \cite{Bena:2018bbd}.

Finding an interpretation of our solutions in the context of 4d black holes is clearly a direction that should be investigated, possibly along the different lines of \cite{Benini:2015eyy,Belin:2019rba,Aniceto:2020saj}. It should be important to clarify  the relation between the number of vacua/holographic central charge and the entropy of these black holes, as advanced around eq.(\ref{4.0}). It would be interesting to understand the role of the freedom in choosing $\hat{h}_4, h_8$ and their implications for black holes.  Similarly, it would be interesting to explore the uses of the formalism developed in \cite{Fedoruk:2015lza}, applied to our particular systems.
Along this line, the possibility of understanding our AdS$_2$ backgrounds as the emergent dynamics in \cite{Anninos:2013nra} is interesting  to explore.
%
\\
We hope to tackle some of these problems in the near future.

\section*{Acknowledgements}  We would like to thank Jerem\'{\i}as Aguilera-Damia,  Dionysios Anninos, Iosif Bena, Panos Betzios, Nikolay Bobev, Alejandra Castro, Diego Correa, Chris Couzens, Giuseppe Dibitetto, Gast\'on Giribet,  Prem Kumar,  Niall Macpherson, Ioannis Papadimitriou, Nicol\`o Petri, Guillermo Silva, David Turton,
 for very useful discussions.
\\
Y.L. and A.R. are partially supported by the Spanish government grant
PGC2018-096894-B-100 and by the Principado de Asturias through the grant FC-GRUPINIDI/
2018/000174. AR is supported by CONACyT-Mexico.

\appendix
\section{AdS$_3$ and AdS$_2$ backgrounds in full generality}\label{appendixgeneral}
 In Section \ref{geometria} we discussed a particular set of solutions in class I of the paper \cite{Lozano:2019emq} and  in Section \ref{A-type} we discussed the T-dual of these ``seed'' backgrounds.
In this appendix we summarise the  general backgrounds
in class I of     \cite{Lozano:2019emq} and perform the T-duality on the AdS$_3$ fibre, generating  AdS$_2$ backgrounds that generalise those of Section \ref{A-type}. 

The Neveu-Schwarz sector of the generic AdS$_3$ backgrounds in \cite{Lozano:2019emq} reads,
\begin{equation}\label{eq:class I backgroundap}
\begin{split}
\text{d}s^2 &= \frac{u}{\sqrt{\hat{h}_4 h_8}} \left( \text{d}s^2_{\text{AdS}_3} + \frac{\hat{h}_4 h_8}{4 \hat{h}_4 h_8 + (u')^2} \text{d}s^2_{\text{S}^2} \right) + \sqrt{\frac{\hat{h}_4}{h_8}} \text{d}s^2_{\text{CY}_2} + \frac{\sqrt{\hat{h}_4 h_8}}{u} \text{d} \rho^2 \, , \\
e^{- \Phi}&= \frac{h_8^{3/4}}{2 \hat{h}_4^{1/4} \sqrt{u}} \sqrt{4 \hat{h}_4 h_8 + (u')^2} \, , \quad H_{3} = \frac{1}{2} \text{d} \bigg( - \rho + \frac{u u'}{4 \hat{h}_4 h_8 + (u')^2} \bigg) \wedge \text{vol}_{\text{S$^2$}}+ \frac{1}{h_8^2} \text{d} \rho \wedge H_2  \, .
\end{split}
\end{equation}
Here the metric is given in the string frame, $\Phi$ is the dilaton and $H_3=\mathrm{d}B_2$ is the NS three-form. 
In the general case the warping function $\hat{h}_4$ has support on $(\rho, \text{CY}_2)$. The RR fluxes are,
 \begin{equation}\label{eq:class I background RRap}
\begin{split}
F_{0} &= h_8' \, , \quad F_{2} =-\frac{1}{h_8}H_2  - \frac{1}{2} \Big( h_8 - \frac{h_8' uu'}{4 h_8 \hat{h}_4+(u')^2} \Big) \text{vol}_{\text{S$^2$}} \, , \\
F_{4} &= -\left( \text{d} \bigg( \frac{uu'}{2 \hat{h}_4} \bigg) + 2 h_8 \text{d} \rho \right) \wedge \text{vol}_{\text{AdS$_3$}}   - \partial_{\rho} \hat{h}_4 \text{vol}_{\text{CY$_2$}}- \frac{h_8}{u} (\hat{\star}_4 \text{d}_4 \hat{h}_4) \wedge \mathrm{d} \rho \\&- \frac{uu'}{2h_8(4 \hat{h}_4 h_8 + (u')^2)} H_2 \wedge \text{vol}_{\text{S$^2$}}   \, ,
\end{split}
\end{equation}
with the higher fluxes related to these as $F_{6} = - \star_{10} F_{4}$, $F_{8} = \star_{10} F_{2}$, $F_{10} = - \star_{10} F_{0}$, and where $\widehat{\star}_4$ is the Hodge dual on the CY$_2$. It was shown in \cite{Lozano:2019emq} that supersymmetry holds whenever,
\begin{equation}\label{supersymmetry conditions}
u'' = 0 \, ,\qquad H_2+\widehat{\star}_4H_2=0\, ,
\end{equation}
which makes $u$ a linear function of $\rho$. 
$H_2$ can be defined in terms of three  functions $g_{1,2,3}$ on CY$_2$, 
\begin{eqnarray} \label{ges}
H_2=g_1(\hat{e}^1\wedge \hat{e}^2-\hat{e}^3\wedge \hat{e}^4)+g_2(\hat{e}^1\wedge \hat{e}^3+\hat{e}^2\wedge \hat{e}^4)+g_3(\hat{e}^1\wedge \hat{e}^4-\hat{e}^2\wedge \hat{e}^3)	,
\end{eqnarray}
where $\widehat{e}^i$ are a canonical vielbein on CY$_2$ (see section 3.1. of \cite{Lozano:2019emq}). Hence, the Bianchi identities of the fluxes impose (away from localised sources),
\begin{eqnarray}
\begin{split}
&h_8''=0\, , \qquad \mathrm{d}H_2=0, \\
&\frac{h_8}{u}\nabla_{\text{CY}_2}^2\hat{h}_4+\partial_\rho^2\hat{h}_4+\frac{2}{h_8^3}(g_1^2+g_2^2+g_3^2)=0.
\end{split}
\end{eqnarray}
In the case when  $H_2$ vanishes and $\widehat{h}_4$  has support on the $\rho$ coordinate only,  we are in the case of the  solutions reviewed in section \ref{geometria}.

We T-dualise the previous backgrounds on the Hopf direction of AdS$_3$ by parametrising it as in \eqref{nara}. Performing T-duality on $\tilde{\psi}$ results in the dual NS sector,
\begin{equation}\label{t dualised background NS appendix}
\begin{split}
\text{d}s^2 &= \frac{u}{\sqrt{\hat{h}_4 h_8}} \left( \frac{1}{4}\text{d}s^2_{\text{AdS}_2} + \frac{\hat{h}_4 h_8}{4 \hat{h}_4 h_8 + (u')^2} \text{d}s^2_{\text{S}^2} \right) + \sqrt{\frac{\hat{h}_4}{h_8}} \text{d}s^2_{\text{CY}_2} + \frac{\sqrt{\hat{h}_4 h_8}}{u} (\text{d} \rho^2 + \text{d} \psi^2 )\, , \\
e^{- 2 \Phi}&= \frac{h_8}{4\hat{h}_4} \big(4 \hat{h}_4 h_8 + (u')^2 \big) \, , \\\quad H_{3} &= \frac{1}{2} \text{d} \bigg( - \rho + \frac{u u'}{4 \hat{h}_4 h_8 + (u')^2} \bigg) \wedge \text{vol}_{\text{S$^2$}}+ \frac{1}{h_8^2} \text{d} \rho \wedge H_2 + \frac{1}{2}\text{vol}_{\text{AdS$_2$}} \wedge \text{d} \psi \, ,
\end{split}
\end{equation}
and the RR sector is,
\begin{equation}\label{t dualised background RR appendix}
\begin{split}
F_{1} &= h_8' \text{d} \psi \, , \\ F_{3} &=  - \frac{1}{2} \Big( h_8 - \frac{h_8' u' u}{4 h_8 \hat{h}_4+(u')^2} \Big) \text{vol}_{\text{S$^2$}} \wedge \text{d} \psi -\frac{1}{h_8}H_2\wedge\text{d} \psi+ \frac{1}{4} \left( \text{d} \bigg( \frac{u'u}{2 \hat{h}_4} \bigg) + 2 h_8 \text{d} \rho \right) \wedge \text{vol}_{\text{AdS$_2$}} \, , \\
F_{5} &=-(1+\star_{10})\left(\partial_{\rho} \hat{h}_4 \text{vol}_{\text{CY$_2$}}+ \frac{h_8}{u} (\hat{\star}_4 \text{d}_4 \hat{h}_4) \wedge \mathrm{d} \rho + \frac{uu'}{2h_8(4 \hat{h}_4 h_8 + (u')^2)} H_2 \wedge \text{vol}_{\text{S$^2$}}  \right) \wedge \mathrm{d} \psi \, ,
\end{split}
\end{equation}
where $F_{7}=- \star_{10} F_{3}=$ and $F_{9}=\star_{10} F_{1}$.

In the case when  $H_2$ vanishes and $\widehat{h}_4$  has support on the $\rho$ coordinate only,  we are in the case of the  solutions constructed in Section \ref{A-type}.

\section{$\mathcal{N} = (0,2)$ and $\mathcal{N} = (0,4)$ theories and Quantum Mechanics}\label{appendix1}

\paragraph{} In this appendix we briefly discuss the QFTs that conjecturally flow in the IR to a strongly coupled CFT. The discussion requires some standard aspects of two dimensional ${\mathcal N} = (0,2) $ and ${\mathcal N} = (0,4)$ supersymmetric theories. As these are well summarised in other works -- see for example \cite{Tong:2014yna,Putrov:2015jpa,Franco:2015tna} -- we shall not give too many details.

$\mathcal{N} = (0,2)$ \textbf{multiplets:} Let us list the field components of the three types of $\mathcal{N} = (0,2)$ multiplets, namely the vector $U$, chiral $\Phi$ and the Fermi $\Psi$ multiplets
\begin{equation}
\boxed{U : (u_{\mu}, \zeta_{-}, D) \, , \quad \quad  \Phi : (\phi, \psi_+) \, , \quad \quad \Psi : (\psi_-, G) \, .}
\end{equation}
The subscript on the fermions refers to their chiralities under the $SO(1,1)$ Lorentz group. $D$ is a real and $G$ a complex auxiliary field.

A \textit{vector} $U$ has the following expansion in superspace\footnote{$\mathcal{N} = (0, 2)$ superspace is parametrised by two real spacetime coordinates, $x_{\pm} = x^0 \pm x^1$, and two complex Grassmann variables $\theta^+$ and $\overline{\theta}^+$ subject to a reality constraint.}
\begin{equation}
U = u_0 - u_1 - 2 i \theta^+ {\bar{\zeta}}_{-} - 2 i {\bar{\theta}}^+ \zeta_{-} + 2 \theta^+ {\bar{\theta}}^+ D \, .
\end{equation}
The corresponding field strength is obtained by means of
\begin{equation}
\Upsilon = [{\overline{\mathcal{D}}}_+, \mathcal{D}_-] = - \zeta_- - i \theta^+ (D - i u_{01}) - i \theta^+ {\bar{\theta}^+} (\mathcal{D}_0+\mathcal{D}_1) \zeta_- \, ,
\end{equation}
where ${\overline{\mathcal{D}}}_+$ and $\mathcal{D}_-$ are the supercovariant gauge derivatives \cite{Witten:1993yc}.
It turns out that $\Upsilon$ is a Fermi multiplet -- it satisfies $\overline{\mathcal{D}}_+ \Upsilon = 0$. We shall give a more precise definition of a Fermi multiplet momentarily.

A \textit{chiral field} $\Phi$ is a superfield that satisfies the following equation
\begin{equation}
{\overline{\mathcal{D}}}_+ \Phi = 0 \, ,
\end{equation}
and therefore expands out in components as
\begin{equation}
\Phi = \phi + \sqrt{2} \theta^+ \psi_+ - i \theta^+ {\bar{\theta}}^+ (D_0 + D_1) \phi \, ,
\end{equation}
where $D_0$ and $D_1$ stand for the time- and space-components of the usual covariant derivative.


A \textit{Fermi superfield} instead satisfies the following equation
\begin{equation}\label{eq:defining fermi multiplet}
{\overline{\mathcal{D}}}_+ \Psi = E (\Phi_i) \, ,
\end{equation}
where $E(\Phi_i)$ is a holomorphic function of the chiral superfields $\Phi_i$. $E$ should be chosen in such a way that it transforms as $\Psi$ under all symmetries. Solving \eqref{eq:defining fermi multiplet} leads to the following expansion for $\Psi$
\begin{equation}
\Psi = \psi_- -  \theta^+ G - i \theta^+ {\bar{\theta}}^+ (D_0 + D_1) \psi_- -  {\bar{\theta}}^+ E(\phi_i) -  \theta^+ {\bar{\theta}}^+ \frac{\partial E}{\partial \phi^i} \psi_{+ i} \, ,
\end{equation}
where $G$ is an auxiliary complex field. The holomorphic function $E$ can be shown to appear in the Lagrangian as a potential term.
There is also another type of superpontential we can consider for $\mathcal{N} = (0,2)$ theories. For each Fermi multiplet $\Psi_a$ we can introduce a holomorphic function $J^a(\Phi_i)$ such that
\begin{equation}
S_J = \int \mathrm{d}^2 x \mathrm{d} \theta^+ \sum_a J^a(\Phi_i) \Psi_a + \text{h.c.} \, .
\end{equation}

It must be stressed that the superpotentials $E$ and $J$ cannot be introduced independently. It turns out that, in order for supersymmetry to be preserved, they have to satisfy the following constraint
\begin{equation}\label{supersymmetric condition on superpotentials}
E \cdot J = \sum_a E_a J^a = 0 \, .
\end{equation}

Let us now move on to listing $\mathcal{N} = (0,4)$ supermultiplets.

$\mathcal{N} = (0,4)$ \textbf{multiplets:} $\,$ $\mathcal{N} = (0,4)$ supermultiplets are usually given in terms of $\mathcal{N} = (0,2)$ supermultiplets, pretty much as in 4 dimensions $\mathcal{N} = 2$ superfields are built from $\mathcal{N} = 1$ superfields. Again, let us list them first.

\begin{center}
\begin{tabular}{ |c|c|c|c| } 
 \hline
 Multiplets & $\mathcal{N} = (0,2)$ building blocks & component fields& $\text{SU}(2)_L \times \text{SU}(2)_R$\\ 
  \hline
 Vector & Vector $+$ Fermi $(U, \Theta)$ & $(u_{\mu}, \zeta_{-}^a, G^A)$& $(1, 1),(2, 2),(3,1)$\\ 
 \hline
 Hyper & Chiral $+$ Chiral $(\Phi, \tilde{\Phi})$ & $(\phi^a, \psi_{+}^b)$& $(2, 1),(1, 2)$\\ 
 \hline
 Twisted hyper & Chiral $+$ Chiral $(\Phi', \tilde{\Phi}')$ & $({\phi'}^a, {\psi'}_{+}^b)$& $(1, 2),(2, 1)$\\
  \hline
Fermi & Fermi $+$ Fermi $(\Gamma, \tilde{\Gamma})$ & $({\psi'}_{-}^a, G^b)$& $(1, 1),(2,2)$\\
  \hline
\end{tabular}
\end{center}

The $\mathcal{N} = (0,4)$ vector multiplet is made of an $\mathcal{N} = (0,2)$ vector multiplet and an adjoint $\mathcal{N} = (0,2)$ Fermi multiplet $\Theta$. The field content is that of a gauge field $u_{\mu}$ and two left-handed fermions $\zeta_{-}^a$, $a = 1, 2$, in addition to a triplet of auxiliary fields $G^A$, $A=1, 2, 3$. The gauge field is a singlet under the $\text{SU}(2)_L \times \text{SU}(2)_R$ R-symmetry while the two fermions transform as $(\mathbf{2}, \mathbf{2})$. The triplet of auxiliary fields transforms as $(\mathbf{3}, \mathbf{1})$ under the R-symmetry.

There are two different types of hypermultiplets, the hypermultiplet and the twisted hypermultiplet. Both of them are formed by two $\mathcal{N} = (0,2)$ chiral multiplets, therefore they both contain two complex scalars ($\phi^a$) and two right-handed fermions ($\psi_{+}^b$). They differ from each other because of the different representations under the R-symmetry group, as we can see from the table above.

If we want to couple the hypermultiplet to the vector multiplet, we should consider the following coupling between the hyper $(\Phi , \tilde{\Phi})$ and the adjoint Fermi field $\Theta$
\begin{equation}
J^{\Theta} = \Phi \tilde{\Phi} \Rightarrow \mathcal{W} = \tilde{\Phi} \Theta \Phi \, .
\end{equation}
On the other hand, coupling a twisted hypermultiplet to the gauge sector requires an E-type of superpotential
\begin{equation}
E_{\Theta} = \Phi' \tilde{\Phi}' \, , 
\end{equation}
with indices in $\Phi' \tilde{\Phi}'$ set to have $E_{\Theta}$ transforming in the adjoint of the gauge group.

Finally, we can have an $\mathcal{N} = (0,4)$ Fermi multiplet, which is made of two $\mathcal{N} = (0,2)$ Fermi multiplets. It contains two left-handed fermions which are singlets of $SU(2)_L \times SU(2)_R$ R-symmetry.
No further coupling between $\Gamma$, $\tilde{\Gamma}$ and $\Theta$ is possible.
\vspace{3pt}

As in the quiver of Figure \ref{figurageneral}, there appear also $\mathcal{N} = (4, 4)$ vector and chiral multiplets, it is worth mentioning how $\mathcal{N} = (4, 4)$ superfields decompose in $\mathcal{N} = (0, 4)$ language.

$\mathcal{N} = (4,4)$ \textbf{multiplets:} There are two types of $\mathcal{N} = (4,4)$ superfields, the vector and the hypermultiplet.

\begin{center}
\begin{tabular}{ |c|c|c| } 
 \hline
 Multiplets & $\mathcal{N} = (0,4)$ building blocks & $\mathcal{N} = (0,2)$ building blocks \\ 
  \hline
 Vector & Vector $+$ Twisted Hyper & $(U, \Theta), (\Sigma, \tilde{\Sigma})$ \\ 
 \hline
 Hyper & Hyper $+$ Fermi & $(\Phi, \tilde{\Phi}), (\Gamma, \tilde{\Gamma})$ \\ 
 \hline
\end{tabular}
\end{center}
The $\mathcal{N}=(4,4)$ vector multipled is comprised of an $\mathcal{N}=(0,4)$ vector multiplet and a $\mathcal{N}=(0,4)$ twisted hypermultiplet. The twisted hypermultiplet is usually denoted as $(\Sigma, \tilde{\Sigma})$. They are coupled to the gauge sector via the E-type potential
\begin{equation}
E_{\Theta} = [\Sigma, \tilde{\Sigma}] \, .
\end{equation}
$\mathcal{N} = (4, 4)$ hypermultiplets are made of an $\mathcal{N} = (0, 4)$ hypermultiplet and an $\mathcal{N} = (4, 4)$ Fermi multiplet, all in all $(\Phi, \tilde{\Phi}), (\Gamma, \tilde{\Gamma})$. As before, $\Phi$ and $\tilde{\Phi}$ are coupled to the gauge sector via
\begin{equation}
\mathcal{W} = \tilde{\Phi} \Theta \Phi \, .
\end{equation}

We conclude this part by stressing out that there are couplings between $\mathcal{N} = (0,4)$ Fermi multiplets $\Gamma$, $\tilde{\Gamma}$, hypermultiplets $\Phi$, $\tilde{\Phi}$ and twisted hypers $\Sigma$, $\tilde{\Sigma}$. They involve both superpotential and E-terms
\begin{equation}
\mathcal{W} = \tilde{\Gamma} \tilde{\Sigma} \Phi + \tilde{\Phi} \tilde{\Sigma} \Gamma \, ,
\end{equation}
and
\begin{equation}
E_{\Gamma} = \Sigma \Phi \, ,\quad E_{\tilde{\Gamma}} = - \tilde{\Phi} \Sigma \, .
\end{equation}
It is easy to see that
\begin{equation}
E \cdot J = \tilde{\Phi} [\Sigma, \tilde{\Sigma}] \Phi + \tilde{\Phi} \tilde{\Sigma} \Sigma \Phi - \tilde{\Phi} \Sigma \tilde{\Sigma} \Phi = 0 \, . 
\end{equation}

\subsection{$\mathcal{N}=4$ Quantum Mechanics}

As we argued in the main text, the $\mathcal{N} = 4$ superconformal quantum mechanics dual to the IIB backgrounds discussed around \eqref{t dualised background NS} and \eqref{t dualised background RR} is given by the dimensional reduction of the CFT in Figure \ref{figurageneral}. Thus, we start with a general discussion on compactification of 2d $\mathcal{N} = (0,4)$ theories. These are usually formulated in terms of $\mathcal{N} = (0,2)$ multiplets, so we start by reducing them first. 

\subsubsection*{$\mathcal{N} = 2$ supersymmetry multiplets}

In $\mathcal{N} = 2$ quantum mechanics we have two real supercharges with an $SO(2)$ R-symmetry. Equivalently, they can be rearranged as two complex supercharges $Q$ and $\overline{Q}$ with a reality constraint, and $U(1)$ R-symmetry. They satisfy the algebra
\begin{equation}
Q^2 = \overline{Q}^2 = 0 \, , \quad \{ Q, \overline{Q} \} = H \, ,
\end{equation}
with $H$ the hamiltonian. Moreover, if we denote by $J$ the R-symmetry generator we have
\begin{equation}
[J, Q] = - Q \, , \quad [J, \overline{Q}] = \overline{Q} \, , \quad [J, H] = 0 \, .
\end{equation}

Let us now see what $\mathcal{N} = 2$ supermultiplets in quantum mechanics are relevant for us. Much of the construction is obtained from the dimensional reduction of 2d $\mathcal{N} = (0, 2)$ reviewed above.

As we have seen in the previous section, the 2d $\mathcal{N} = (0,2)$ vector multiplet consists of a two-dimensional gauge field $u_{\mu}$, a left-handed (complex) fermionic field $\zeta_{-}$ and a real auxiliary field $D$. They are all valued in the adjoint representation of the corresponding gauge group. In the following we will just set $\zeta_{-} \equiv \zeta$, as there is no chirality in 1d. After reduction, we have $u_{\mu} = (u_t, \sigma)$, where $u_{t}$ is the one dimensional gauge field and $\sigma$ a real scalar. The supersymmetric kinetic term for an $\mathcal{N} = 2$ vector multiplet in quantum mechanics is
\begin{equation}
L_{\text{vector}} = \frac{1}{2 g^2} \text{tr} \left[ (D_t \sigma)^2 + i \bar{\zeta} D_{t}^{(+)} \zeta + D^2 \right] \, ,
\end{equation}
where $D_{t}^{(\pm)} = D_t \pm i \sigma$ and $D_t$ is the usual covariant derivative $D_{t} = \partial_t + i u_t$ for fields in a generic representation of the gauge group.

A 2d $\mathcal{N} = (0,2)$ chiral multiplet consists of a complex scalar boson $\phi$ and a right-handed (complex) fermionic field $\psi_{+}$ in some unitary representation of the gauge group. As before, we will only be concerned with the fundamental and adjoint representations. Again, in going down to 1d we will drop the sub-index. The supersymmetric kinetic term for an $\mathcal{N} = 2$ chiral multiplet in quantum mechanics reads
\begin{equation}
L_{\text{chiral}} =  D_t \bar{\phi} D_t \phi + i \bar{\psi} D_{t}^{(-)} \psi + \bar{\phi} (D - \sigma^2) \phi - i \sqrt{2} \bar{\phi} \zeta \psi + i \sqrt{2} \bar{\psi} \bar{\zeta} \phi \, .
\end{equation}

A 2d $\mathcal{N} = (0,2)$ Fermi multiplet consists on a left-handed (complex) fermion $\psi_{-}$ and an auxiliary field $G$. In the following, we will make the identification $\psi_{-} \equiv \eta$ and $\psi_{+, i} \equiv \psi_{i}$ if $(\phi_i, \psi_{+, i})$ is a chiral multiplet. The Lagrangian for a generic Fermi multiplet reads
\begin{equation}
L_{\text{fermi}} = i \bar{\eta} D_t^{(+)} \eta + |G^2| - |E(\phi_i)|^2 - \bar{\eta} \frac{\partial E}{\partial \phi_i} \psi_i - \bar{\psi_i} \frac{\partial E}{\partial \bar{\phi}_i} \eta \, .
\end{equation}
In addition to the $E$-term potentials it is possible, for each Fermi multiplet $\Psi_a$, to introduce a holomorphic function $J^a(\Phi_i)$ which gives rise to interactions of the form
\begin{equation}
L_{J} = G^a J_a(\phi_i) + \sum_i \eta_a \frac{\partial J^a}{\partial \phi^i} \psi_i + \text{h.c.} \, .
\end{equation}
As remarked already, the superpotentials $E$ and $J$ cannot be introduced independently. In order for supersymmetry to be preserved, they must satisfy $\sum_a E_a J^a = 0$.

\subsubsection*{$\mathcal{N}=4$ supersymmetric systems}

The $\mathcal{N} = 4$ supermultiplets that are relevant to our construction are given just by dimensional reduction of $\mathcal{N} = (0,4)$ and $\mathcal{N} = (4,4)$ supermultiplets. Two-dimensional $\mathcal{N} = (0,4)$ and $\mathcal{N} = (4,4)$ supermultiplets are given in terms of $\mathcal{N} = (0,2)$ multiplets, according to the discussion in the previous section, summarised in the two tables above. The dimensional reduction of the 2d theory depicted in Figure \ref{figurageneral} is then readily done according to the rules above. In particular, a two-dimensional gauge field always reduces to one-component gauge field plus a scalar in one dimension. Scalars and fermions remain untouched. In the case of the fermions, this is due to the fact that in both one and two dimensions the minimal spinor representation is one-dimensional. Supersymmetric interactions for the UV Lagrangian can be added as long as the condition \eqref{supersymmetric condition on superpotentials} is satisfied. See for instance \cite{Tong:2014yna}.

Before ending this section, let us give one remark about the R-symmetry of the IR theory.

\subsubsection*{R-symmetry}

\paragraph{} The R-symmetry group of supersymmetric $\mathcal{N} = 4$ quantum mechanics is $\text{SO}(4) = \text{SU}(2) \times \text{SU}(2)$. As we flow to the IR and hit a fixed point, given that it exists, we should find that our quantum mechanics realises some classified superconformal algebra. When $\mathcal{N} = 4$, we have essentially two possibilities: $\mathfrak{d}(2, 1; \alpha)$, with two $\mathfrak{su}(2)$'s R-symmetries, or $\mathfrak{su}(1, 1|2)$, with one $\mathfrak{su}(2)$ only.

The $\mathfrak{d}(2, 1; \alpha)$ global algebra is often referred to as \textit{large} superconformal algebra and $\alpha$ is a parameter which parametrises the relative strength of the two Kac-Moody levels, $k_{-}$ and $k_{+}$ of the SU$(2)$ R-symmetries. Given that we have only one SU$(2)$ (realised geometrically on the S$^2$) and given also that in the parent $\text{AdS}_3 \times \text{S}^2$ backgrounds supersymmetries were in the $(\mathbf{1, 2;2})$ of $\text{SL}(2, \mathbb{R}) \times \text{SL}(2, \mathbb{R}) \times \text{SU}(2)_R$, we are naturally led to the conclusion that the (global) superalgebra realised by our backgrounds and the dual field theories is the $\mathfrak{su}(1, 1|2)$ superalgebra\footnote{The $\mathfrak{su}(1, 1|2)$ is also realised by taking the limit $\alpha \rightarrow \infty$ in the $\mathfrak{d}(2, 1; \alpha)$ algebra.}. Also, superalgebras in one and two dimensions are closely related -- each chiral sector of a 2d SCFT provides a superalgebra and its realisation for a 1d superconformal QM -- and this makes it possible to identify central charges in 1d and 2d \cite{Balasubramanian:2009bg}.

\section{Singularity structure at the ends of the $\rho$-interval}\label{rhointerval}
In this appendix we study the asymptotic behaviour of the  backgrounds in eq.(\ref{t dualised background NS}), for defining functions $\hat{h}_4, h_8$ given by eqs.(\ref{profileh4sp})-(\ref{profileh8sp}). Other possible ways of bounding the space can be considered following \cite{Lozano:2019emq,Lozano:2019zvg}.
We distinguish two cases:
\begin{itemize}
\item{$u= c_u \rho$: At $\rho=0$, we find a regular background. In turn,
at the end of the $\rho$-interval (which we denote by $\rho=2\pi(P+1)-x$ for small $x$) we find a metric and dilaton that behave as,
\begin{equation}
\mathrm{d}s^2\sim \frac{1}{x} \mathrm{d}s^2_{\text{AdS}_2} + x(\mathrm{d}x^2+  \text{d}\psi^2 + \mathrm{d}s^2_{\text{S}^2} ) + \mathrm{d}s^2_{\textrm{CY$_2$}},\;\;\;\; e^{-2\Phi}\sim 1.\label{forappA}
\end{equation}
This is a superposition of O1 and O5 planes, extended on AdS$_2$ (and smeared on the $\text{CY}_2$) and AdS$_2$ $\times $ CY$_2$ respectively (see for example
around equation (3.38) of the paper \cite{Lozano:2019emq}).
 }
\item{$u=u_0$: At both ends of the interval,  the metric and dilaton asymptote similarly. The expansion of the background at both ends  is,
\begin{equation}
\mathrm{d}s^2\sim \frac{1}{\rho}(\mathrm{d}s^2_{\text{AdS}_2} + \mathrm{d}s^2_{\text{S}^2}) + \rho(\text{d}\rho^2+ \text{d}\psi^2)  + \mathrm{d}s^2_{\textrm{CY$_2$}},\;\;\;\; e^{-2\Phi}\sim \rho^2.\label{forappA2}
\end{equation}
This indicates the superposition of O3 and O7 planes, extended on AdS$_2 \times \text{S}^2$  (and smeared on the $\text{CY}_2$) and AdS$_2\times \text{S}^2\times $ CY$_2$ respectively.
}
\end{itemize}
In both cases we find that in approaching the end of the interval, the $\psi$-cycle becomes of vanishing size. T-dualising in this direction we recover the seed backgrounds discussed in Section \ref{geometria}.

Analysing the  volume of  the compact submanifolds of the solutions in eqs.(\ref{profileh4sp})-(\ref{profileh8sp}) with $u'=0$, we run into the possibility that some of these submanifolds have infinite size. However, in spite of a divergent warp factor, the ``stringy size'' of the submanifold is actually finite or vanishing at the ends of the space. The finite stringy-volume case does not pose any problem in interpreting a D-brane wrapping such cycle. The case in which the cycle shrinks may suggest an interpretation of the singularity in terms of new massless degrees of freedom (branes wrapping the shrinking cycles) that the supergravity solution is not encoding. 

\begin{figure}
\centering
\includegraphics[scale=0.55]{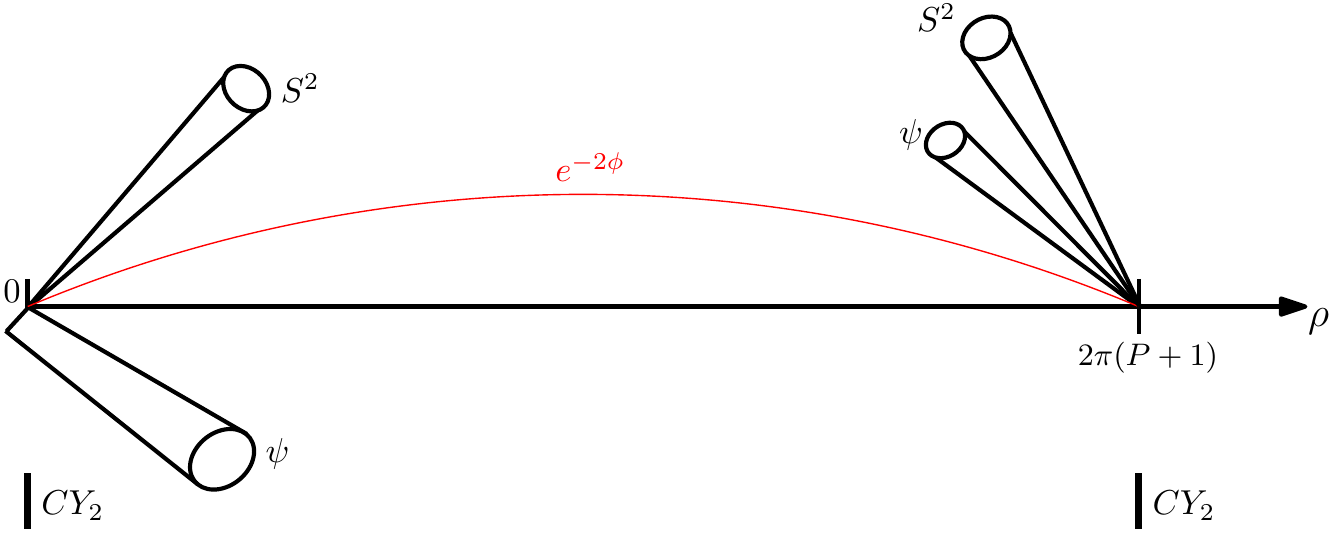}
\caption{Behaviour of the solutions at both ends of the $\rho$-interval for $u= u_0\rho$. The $\text{S}^2$ vanishes, while the $\text{S}^1_\psi$ is finite at $\rho=0$ but shrinks to zero size at $\rho=2\pi (P+1)$. The CY$_2$ has finite size at both ends.}
\label{figuraA1}
\end{figure}

\begin{figure}
\centering
\includegraphics[scale=0.55]{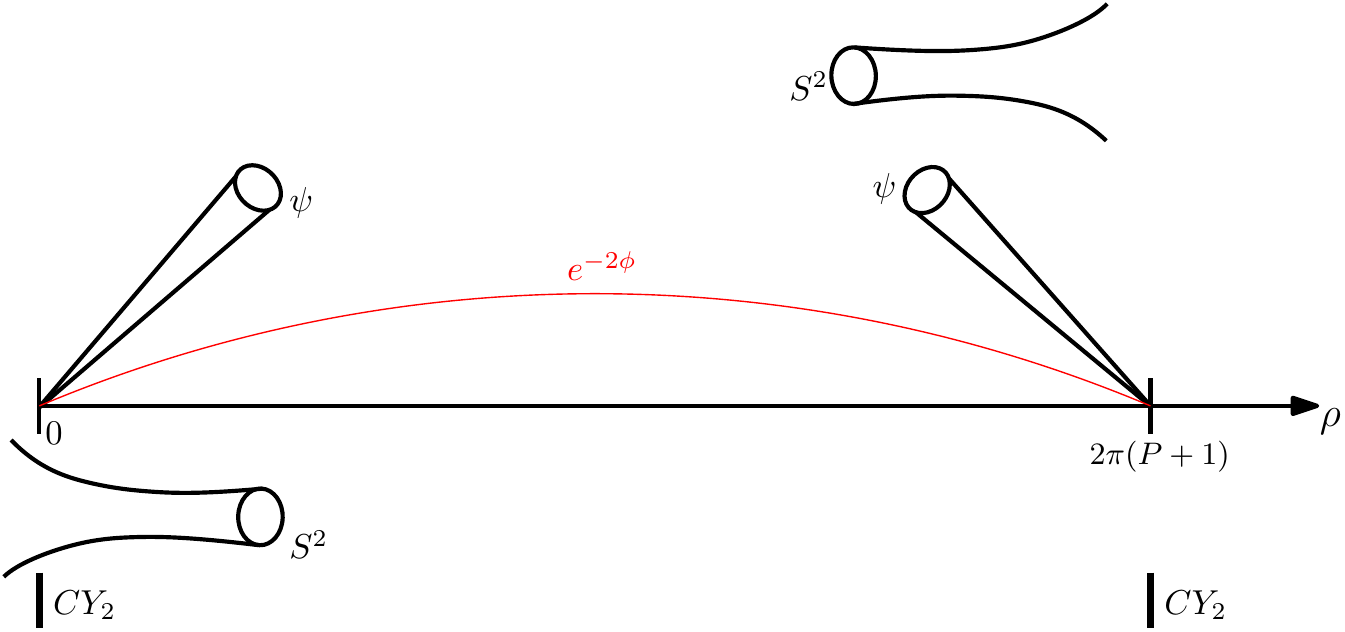}
\caption{Behaviour of the solutions at both ends of the $\rho$-interval for $u= u_0$. The $\text{S}^2$ diverges while the $\text{S}^1_\psi$ shrinks at both ends. The CY$_2$ remains finite.}
\label{figuraA2}
\end{figure}

For the case in eqs.(\ref{profileh4sp})-(\ref{profileh8sp}), with $u= u_0\rho$ and hence non-vanishing $u'$, the background is smooth at $\rho=0$, but it presents a singularity at $\rho=2\pi(P+1)$. We analysed this singularity around eq.(\ref{forappA}). A pictorial view of this background is given in Figure \ref{figuraA1}.
In the case in which $u'=0$, the asymptotic behaviour  given in eq.(\ref{forappA2}) is sketched 
in Figure \ref{figuraA2}. 

In spite of the two sphere having divergent volume, we find that the stringy volume of the $\text{S}^2$ calculated as,
\begin{equation}
V_{s}[\text{S}^2]= \int \text{vol}_{\text{S}^2}\; e^{-\Phi}\sqrt{\det[g+B]}=2\pi \sqrt{h_8^2 \left(\frac{u^2}{16 \hat{h}_4 h_8} +\frac{(\rho-2\pi k)^2}{4}\right)} ,
\end{equation}
is finite for $\rho=0$ and $\rho=2\pi (P+1)$. A brane wrapped on $\text{S}^2$ will then have finite energy and will not pose problems when considering its backreaction.

\section{Holographic calculation of QFT observables}\label{othercalculations}

In this appendix we discuss the holographic
calculation of various field theoretical observables of the strongly coupled quantum mechanics. We focus on Wilson loops, baryon vertices and gauge couplings.

\subsection{Wilson Loops }
As we mentioned in the main text we expect that our conformal quantum mechanics are related to the more general theories describing line defects inside five dimensional ${\cal N}=2$ SCFTs, studied in \cite{Assel:2018rcw,Assel:2019iae}. This opens the possibility that the VEV of a Wilson (or 't Hooft) line can be exactly computed using localisation, along the lines described in \cite{Assel:2019iae,Aguilera-Damia:2017znn}. Here we discuss the holographic calculation of a particular Wilson line that can potentially be checked with some exact methods.

Consider a fundamental string extended on AdS$_2$, parametrised with coordinates $(t,r)$ as in eq.(\ref{nara}). The string has a profile $\rho=\rho(r)$. The induced metric and  NS-NS 2-form field, as well as the action for the string, are obtained from eqs.(\ref{nara})-(\ref{t dualised background NS}). They read,
\begin{eqnarray}
& & \mathrm{d}s_{ind}^2= \frac{u}{4\sqrt{\hat{h}_4 h_8} }(-\mathrm{d}t^2 \cosh^2 r + \mathrm{d}r^2) + \frac{\sqrt{\hat{h}_4 h_8} }{u} \rho'^2 \mathrm{d}r^2,\nonumber\\
& & B_2= \frac{\psi_0}{2} \cosh r \mathrm{d}t\wedge \mathrm{d}r,\nonumber\\
& & S_{F1}= \frac{1}{2\pi} \int \mathrm{d}t \mathrm{d}r \cosh r \left[   \frac{u}{4\sqrt{\hat{h}_4 h_8}   }\sqrt{1+ \frac{4 \hat{h}_4 h_8}{u^2}   \rho'^2}  -\frac{\psi_0}{2}\right].
\end{eqnarray}
We solve the equations of motion for this probe string if
\begin{equation}
\partial_\rho\Bigl( \frac{u}{ 4\sqrt{\hat{h}_4 h_8 } }\Bigr)=0, \qquad u=u_0\rho,\qquad \hat{h}_4= \frac{\beta}{2\pi}\rho, \qquad h_8=\frac{\nu}{2\pi}\rho.\label{popa}
\end{equation}
The solution in eq.(\ref{popa}) implies that the string is sitting close to the beginning of a generic quiver, for the functions $\hat{h}_4,h_8$ in eqs.(\ref{profileh4sp})-(\ref{profileh8sp}).
The on-shell action for this string is,
\begin{equation}
S_{on-shell}= \frac{1}{2\pi}\int \mathrm{d}t \mathrm{d}r \cosh r \left(\frac{\pi u_0}{2\sqrt{\nu\beta}} -\frac{\psi_0}{2}\right)=\frac{1}{2\pi}\left(\frac{\pi u_0}{2\sqrt{\nu\beta}} -\frac{\psi_0}{2} \right) \textrm{Vol}_{\textrm{AdS}_2}.
\end{equation}
This is the quantity that we associate with the expected value of this particular Wilson loop.

There is another solution with constant profiles
 $u\sim h_8\sim \hat{h}_4\sim 1$.  This solution does not fall within the analysis of this paper. Instead, it can be thought of as the reduction and T-dual, along the fibration in $\text{S}^3$,  of the background AdS$_3\times \text{S}^3\times K3$.
Other interesting defect-like operator, the baryonic vertex, is discussed below.

\subsection{Baryonic Vertex}\label{baryonvertex}
We study here a couple of probe branes that we can identify with baryonic vertices.
As in the paper \cite{Witten:1998xy}, there will be an integer number of fundamental strings ending on them and their tension will be of order $T_{bar}\sim 1/g_s$.

We consider first  the gauge groups that come from D5 branes in the interval $[2\pi k, 2\pi(k+1)]$, for which the number of branes is $\mu_k$. For this gauge group, we consider a probe 
consisting on a D3 brane extended along $[t,\text{S}^2,\psi]$ at some fixed value $\rho=2\pi k$. The induced metric, NS-NS B-field and BIWZ action read,
\begin{eqnarray}
& & \mathrm{d}s_{ind,\text{D3}}^2= -\frac{u}{4\sqrt{\hat{h}_4 h_8}}\cosh^2 r \mathrm{d}t^2 + \frac{u\sqrt{\hat{h}_4 h_8}}{\Delta} \mathrm{d}s^2_{\text{S}^2} + \frac{\sqrt{\hat{h}_4 h_8}}{u} \text{d}\psi^2,\;\;\;\Delta= 4 \hat{h}_4 h_8 + u'^2, \nonumber\\
& & B_2= \frac{1}{2\Delta}\left[(2\pi k-\rho) \Delta + u u'\right]\text{vol}_{S^2}-\frac{\sinh r}{2} \mathrm{d}t\wedge \text{d}\psi,\label{d3probe}\\
& & S_{BI}= T_{\text{D3}}\int  e^{-\Phi}\sqrt{\det[g+B]}\;\mathrm{d}t\wedge \text{d}\psi\wedge\text{vol}_{\text{S}^2}= T_{\text{D3}} \left(\frac{u}{8}\sqrt{\frac{h_8}{\hat{h}_4}} \right)|_{\rho=2\pi k} (8\pi^2)\int \mathrm{d}t,\nonumber\\
& & S_{WZ}= T_{\text{D3}}\int C_4 + f_2\wedge C_2=- T_{\text{D3}}\int_t a_t\int_{\text{S}^2\times \text{S}^1_\psi} \hat{F}_3= \frac{\mu_k}{2\pi}\int_t a_t .\nonumber
\end{eqnarray}
We used that $f_2= da_1$. We see that the mass of this particle is given by $M_{bar}=\pi^2 T_{\text{D3}} \left(u\sqrt{\frac{h_8}{\hat{h}_4}} \right)|_{\rho=2\pi k}$. We also observe that $\mu_k$ strings must end on it, to cancel the charge of the object on $\text{S}^2\times\text{S}^1_\psi$. This is the baryonic vertex for the gauge group U$(\mu_k)$. 

With a probe D7 brane extended in $[t, \textrm{CY$_2$}, \text{S}^2,\psi]$ at some fixed value of $\rho=2\pi k$, we find analogously,
\begin{eqnarray}
& & \mathrm{d}s_{ind,\;\text{D7}}^2= -\frac{u}{4\sqrt{\hat{h}_4 h_8}}\cosh^2 r \mathrm{d}t^2 + \frac{u\sqrt{\hat{h}_4 h_8}}{\Delta} \mathrm{d}s^2_{\text{S}^2} + \sqrt{\frac{\hat{h}_4}{h_8}} \mathrm{d}s^2_{\text{CY}_2}+ \frac{\sqrt{\hat{h}_4 h_8}}{u} \text{d}\psi^2,\nonumber\\
& & \Delta= 4 \hat{h}_4 h_8 + u'^2,\;\;\;B_2= \frac{1}{2\Delta}\left[(2\pi k-\rho) \Delta + u u'\right]\text{vol}_{S^2}-\frac{\sinh r}{2} \mathrm{d}t\wedge \text{d}\psi,\label{d7probe}\\
& & S_{BI}= T_{\text{D7}}\int e^{-\Phi}\sqrt{\det[g+B]}\; \mathrm{d}t\wedge \text{d}\psi\wedge\text{vol}_{\textrm{S$^2$}} \wedge\text{vol}_{\textrm{CY$_2$}}= \nonumber\\
&&=T_{\text{D7}}\;\text{Vol}_{\textrm{CY$_2$}} \text{Vol}_ {\text{S}^2} \text{Vol}_{\psi} \Bigl(\frac{u}{8}\sqrt{\frac{\hat{h}_4}{{h}_8}}\Bigr)\bigr\rvert_{\rho=2\pi k}\int \mathrm{d}t,\nonumber\\
& & S_{WZ}= T_{\text{D7}}\int C_8 + f_2\wedge C_6= -T_{\text{D7}}\int_t a_t \int_{\text{S}^2\times \textrm{CY$_2$}\times \text{S}^1_\psi} \hat{F}_7= \frac{\alpha_k}{2\pi}\int_t a_t .\nonumber
\end{eqnarray}
In this case we find that the mass of the particle is $M=T_{\text{D7}}\;\text{Vol}_{\textrm{CY$_2$}} \text{Vol}_ {\text{S}^2} \text{Vol}_{\psi}\left(\frac{u}{8}\sqrt{\frac{\hat{h}_4}{ h_8}}  \right)\rvert_{\rho=2\pi k}$, and that there should be $\alpha_k$ fundamental strings ending on it.

\subsection{Gauge Couplings}
%
 The gauge coupling of each node can also be computed holographically.
We follow a prescription that works in higher dimensional systems.  For $\alpha_k$ gauge groups we  study the action of D1 branes extending in $[t,\rho]$ with $\rho\in [2\pi k,2\pi (k+1)]$ (at fixed values of the other coordinates). For the $\mu_k$ gauge groups, we study D5 branes that extend on $[t, \rho, \text{CY}_2]$ with $\rho\in [2\pi k,2\pi (k+1)]$, also at fixed values of all other coordinates.

We compute the Born-Infeld and Wess-Zumino actions of these branes. We associate the coefficient of the BI parts with the gauge couplings.
For the D1 brane probe we consider here, for which the gauge field is taken to be zero and for which the NS two form has zero pull-back on the brane worldvolume, the induced metric, Born-Infeld-Wess-Zumino action and gauge coupling $g_{YM,1}^2$ read,
\begin{eqnarray}
& & S_{BIWZ}= T_{\text{D1}}\int \mathrm{d}t \text{d}\rho \, e^{-\Phi}\sqrt{-\det[g_{ind}]} - T_{\text{D1}}\int C_2\label{actionD1}\\
& & \mathrm{d}s_{ind,\text{D1}}^2= -\frac{u}{4\sqrt{\hat{h}_4 h_8}}\cosh^2 r_0 \mathrm{d}t^2+\frac{\sqrt{\hat{h}_4 h_8}}{u} \text{d}\rho^2, \;\;\; C_2= \frac{\sinh(r_0)}{4}\left( \partial_\rho\left(\frac{uu'}{2\hat{h}_4}\right) + 2 h_8\right) \text{d}\rho\wedge \mathrm{d}t,\nonumber\\
& & S_{BI}= -T_{\text{D1}}\int_{2\pi k}^{2\pi (k+1)} \text{d}\rho\,  \sqrt{ \frac{h_8}{\hat{h}_4}  (4 \hat{h}_4h_8+ u'^2) }  \int \mathrm{d}t\, \frac{\cosh r_0}{4},\nonumber\\
& &  S_{WZ}= T_{\text{D1}}\int_{2\pi k}^{2\pi(k+1)} \text{d}\rho\left(\partial_\rho\left(\frac{uu'}{\hat{h}_4} \right) + 2 h_8 \right) \int \mathrm{d}t\,\frac{\sinh r_0}{4},\nonumber
\end{eqnarray}
Notice that this probe D1 brane becomes extremal (its tension equals its charge) when $u'=0$ and when the brane is placed near the boundary of AdS$_2$ (that is, $r_0\to\infty$). Probably under these circumstances the branes are calibrated.  We can define the gauge coupling from the coefficient of the Born-Infeld term. We find for $\alpha_k$  gauge groups,
\begin{equation}
\frac{1}{g_{YM,1}^2[k,k+1]}= \frac{1}{2\pi}\int_{2\pi k}^{2\pi(k+1)} h_8 \text{d}\rho= \frac{2\mu_k+\nu_k}{2}.
\end{equation}
Notice that this coupling is dimensionless. 

Similarly for the $\mu_k$ 
gauge groups we find, using a D5 brane in $[t,\rho, \text{CY}_2]$ (at fixed values for all other coordinates),
\begin{eqnarray}
& & \mathrm{d}s_{ind,\text{D5}}^2= -\frac{u}{4\sqrt{\hat{h}_4 h_8}}\cosh^2 r_0 \mathrm{d}t^2+\frac{\sqrt{\hat{h}_4 h_8}}{u} \text{d}\rho^2 + \sqrt{\frac{\hat{h}_4}{h_8}} \text{d}s^2_{\text{CY}_2} ,\\
& & C_6= \left(  \frac{4 \hat{h}_4 h_8^2 - u u' h_8' + h_8 (u')^2}{8 h_8^2} \right)\sinh r_0 \text{d}\rho\wedge \mathrm{d}t\wedge  \text{vol}_{\text{CY$_2$}}, \nonumber\\
& &S_{BI} =-T_{\text{D5}} {\text{Vol}}_{\text{CY$_2$}} \int_{2\pi k}^{2\pi (k+1)}  \text{d}\rho\, \sqrt{ \frac{\hat{h}_4}{h_8}  (4 \hat{h}_4h_8+ u'^2) }  \int \mathrm{d}t~ \frac{\cosh r_0}{4},\nonumber\\
& & S_{WZ}=T_{\text{D5}} {\text{Vol}}_{\text{CY$_2$}} \int_{2\pi k}^{2\pi (k+1)}  \text{d}\rho\,  \left(  \frac{4 \hat{h}_4 h_8^2 - u u' h_8' + h_8 (u')^2}{8 h_8^2} \right)\int \mathrm{d}t ~{\sinh r_0} ,\nonumber\\
& & \frac{1}{g_{YM,2}^2[k,k+1]} =  \frac{1}{2\pi}\int_{2\pi k}^{2\pi(k+1)} h_4 \text{d}\rho= \frac{2\alpha_k+\beta_k}{2}.\nonumber
\end{eqnarray}
In the last line we observe that this particular D5 brane probe is extremal for the solutions with $u'=0$ and at $r_0\to\infty$.
\\


\begin{thebibliography}{99}


\bibitem{Maldacena:1997re} 
  J.~M.~Maldacena,
  ``The Large N limit of superconformal field theories and supergravity,''
  Int.\ J.\ Theor.\ Phys.\  {\bf 38}, 1113 (1999)
  [Adv.\ Theor.\ Math.\ Phys.\  {\bf 2}, 231 (1998)]
  [hep-th/9711200].


\bibitem{Gaiotto:2009we} 
  D.~Gaiotto,
  ``N=2 dualities,''
  JHEP {\bf 1208}, 034 (2012)
  [arXiv:0904.2715 [hep-th]].

\bibitem{Gaiotto:2009gz} 
  D.~Gaiotto and J.~Maldacena,
  ``The Gravity duals of N=2 superconformal field theories,''
  JHEP {\bf 1210}, 189 (2012)
  [arXiv:0904.4466 [hep-th]].

\bibitem{ReidEdwards:2010qs} 
  R.~A.~Reid-Edwards and B.~Stefanski, jr.,
  ``On Type IIA geometries dual to N = 2 SCFTs,''
  Nucl.\ Phys.\ B {\bf 849}, 549 (2011)
  [arXiv:1011.0216 [hep-th]].
  
\bibitem{Aharony:2012tz} 
  O.~Aharony, L.~Berdichevsky and M.~Berkooz,
  ``4d N=2 superconformal linear quivers with type IIA duals,''
  JHEP {\bf 1208}, 131 (2012)
  [arXiv:1206.5916 [hep-th]].

\bibitem{Lozano:2016kum}
Y.~Lozano and C.~Nunez,
``Field theory aspects of non-Abelian T-duality and $ \mathcal{N}  =$ 2 linear quivers,''
JHEP \textbf{05}, 107 (2016)
[arXiv:1603.04440 [hep-th]].

\bibitem{Nunez:2018qcj} 
  C.~Nunez, D.~Roychowdhury and D.~C.~Thompson,
  ``Integrability and non-integrability in $ \mathcal{N}=2 $ SCFTs and their holographic backgrounds,''
  JHEP {\bf 1807}, 044 (2018)
  [arXiv:1804.08621 [hep-th]].
  C.~Nunez, D.~Roychowdhury, S.~Speziali and S.~Zacarias,
  ``Holographic Aspects of Four Dimensional ${\cal N }=2$ SCFTs and their Marginal Deformations,''
  Nucl.\ Phys.\ B {\bf 943}, 114617 (2019)
  [arXiv:1901.02888 [hep-th]].
%
\bibitem{Bah:2019jts} 
  I.~Bah, F.~Bonetti, R.~Minasian and E.~Nardoni,
  ``Anomaly Inflow for M5-branes on Punctured Riemann Surfaces,''
  JHEP {\bf 1906}, 123 (2019)
  [arXiv:1904.07250 [hep-th]].
I.~Bah, F.~Bonetti, R.~Minasian and P.~Weck,
[arXiv:2002.10466 [hep-th]].

\bibitem{Brandhuber:1999np}
A.~Brandhuber and Y.~Oz,
``The D-4 - D-8 brane system and five-dimensional fixed points,''
Phys. Lett. B \textbf{460} (1999), 307-312
[arXiv:hep-th/9905148 [hep-th]].

\bibitem{Bergman:2012kr}
O.~Bergman and D.~Rodriguez-Gomez,
``5d quivers and their AdS(6) duals,''
JHEP \textbf{07} (2012), 171
[arXiv:1206.3503 [hep-th]].

\bibitem{Lozano:2012au}
Y.~Lozano, E.~\'O Colg\'ain, D.~Rodr\'\i{}guez-G\'omez and K.~Sfetsos,
``Supersymmetric $AdS_6$ via T Duality,''
Phys. Rev. Lett. \textbf{110} (2013) no.23, 231601
[arXiv:1212.1043 [hep-th]].

\bibitem{Lozano:2013oma}
Y.~Lozano, E.~\'O Colg\'ain and D.~Rodr\'\i{}guez-G\'omez,
``Hints of 5d Fixed Point Theories from Non-Abelian T-duality,''
JHEP \textbf{05} (2014), 009
[arXiv:1311.4842 [hep-th]].

\bibitem{DHoker:2016ujz} 
  E.~D'Hoker, M.~Gutperle, A.~Karch and C.~F.~Uhlemann,
  ``Warped $AdS_6\times S^2$ in Type IIB supergravity I: Local solutions,''
  JHEP {\bf 1608}, 046 (2016)
  [arXiv:1606.01254 [hep-th]].
 E.~D'Hoker, M.~Gutperle and C.~F.~Uhlemann,
  ``Holographic duals for five-dimensional superconformal quantum field theories,''
  Phys.\ Rev.\ Lett.\  {\bf 118}, no. 10, 101601 (2017)
  [arXiv:1611.09411 [hep-th]].
  E.~D'Hoker, M.~Gutperle and C.~F.~Uhlemann,
  ``Warped $AdS_6\times S^2$ in Type IIB supergravity II: Global solutions and five-brane webs,''
  JHEP {\bf 1705}, 131 (2017)
  [arXiv:1703.08186 [hep-th]].
  M.~Gutperle, C.~Marasinou, A.~Trivella and C.~F.~Uhlemann,
``Entanglement entropy vs. free energy in IIB supergravity duals for 5d SCFTs,''
JHEP \textbf{09} (2017), 125
[arXiv:1705.01561 [hep-th]].
  E.~D'Hoker, M.~Gutperle and C.~F.~Uhlemann,
``Warped $AdS_6\times S^2$ in Type IIB supergravity III: Global solutions with seven-branes,''
JHEP \textbf{11} (2017), 200
[arXiv:1706.00433 [hep-th]].
  
\bibitem{Gutperle:2018vdd} 
  M.~Gutperle, A.~Trivella and C.~F.~Uhlemann,
  ``Type IIB 7-branes in warped AdS$_{6}$: partition functions, brane webs and probe limit,''
  JHEP {\bf 1804}, 135 (2018)
  [arXiv:1802.07274 [hep-th]].
M.~Fluder and C.~F.~Uhlemann,
  ``Precision Test of AdS$_6$/CFT$_5$ in Type IIB String Theory,''
  Phys.\ Rev.\ Lett.\  {\bf 121}, no. 17, 171603 (2018)
  [arXiv:1806.08374 [hep-th]].

\bibitem{Bergman:2018hin} 
  O.~Bergman, D.~Rodriguez-Gomez and C.~F.~Uhlemann,
  ``Testing AdS$_{6}$/CFT$_{5}$ in Type IIB with stringy operators,''
  JHEP {\bf 1808}, 127 (2018)
  [arXiv:1806.07898 [hep-th]].
 
\bibitem{Lozano:2018pcp}
Y.~Lozano, N.~T.~Macpherson and J.~Montero,
``AdS$_{6}$ T-duals and type IIB AdS$_{6} \times$ S$^{2}$ geometries with 7-branes,''
JHEP \textbf{01} (2019), 116
[arXiv:1810.08093 [hep-th]].
  
\bibitem{Uhlemann:2019ypp}
  C.~F.~Uhlemann,
  ``Exact results for 5d SCFTs of long quiver type,''
  arXiv:1909.01369 [hep-th].
  C.~F.~Uhlemann,
``Wilson loops in 5d long quiver gauge theories,''
[arXiv:2006.01142 [hep-th]].



%
\bibitem{Apruzzi:2015wna}
  F.~Apruzzi, M.~Fazzi, A.~Passias, A.~Rota and A.~Tomasiello,
  ``Six-Dimensional Superconformal Theories and their Compactifications from Type IIA Supergravity,''
  Phys.\ Rev.\ Lett.\  {\bf 115} (2015) no.6,  061601
  [arXiv:1502.06616 [hep-th]].


\bibitem{Apruzzi:2013yva}
  F.~Apruzzi, M.~Fazzi, D.~Rosa and A.~Tomasiello,
  ``All AdS$_7$ solutions of type II supergravity,''
  JHEP {\bf 1404} (2014) 064
  [arXiv:1309.2949 [hep-th]].


\bibitem{Gaiotto:2014lca}
  D.~Gaiotto and A.~Tomasiello,
  ``Holography for (1,0) theories in six dimensions,''
  JHEP {\bf 1412} (2014) 003
  [arXiv:1404.0711 [hep-th]].

\bibitem{Cremonesi:2015bld}
  S.~Cremonesi and A.~Tomasiello,
  ``6d holographic anomaly match as a continuum limit,''
  JHEP {\bf 1605} (2016) 031
  [arXiv:1512.02225 [hep-th]].

\bibitem{Nunez:2018ags}
  C.~Nunez, J.~M.~Penin, D.~Roychowdhury and J.~Van Gorsel,
  ``The non-Integrability of Strings in Massive Type IIA and their Holographic duals,''
  JHEP {\bf 1806} (2018) 078
  [arXiv:1802.04269 [hep-th]].
  K.~Filippas, C.~Nunez and J.~Van Gorsel,
  ``Integrability and holographic aspects of six-dimensional $ \mathcal{N}=\left(1,\ 0\right) $ superconformal field theories,''
  JHEP {\bf 1906}, 069 (2019)
  [arXiv:1901.08598 [hep-th]].

\bibitem{Bergman:2020bvi}
O.~Bergman, M.~Fazzi, D.~Rodriguez-Gomez and A.~Tomasiello,
``Charges and holography in 6d (1,0) theories,''
[arXiv:2002.04036 [hep-th]].

\bibitem{Brunner:1997gf}
  I.~Brunner and A.~Karch,
  ``Branes at orbifolds versus Hanany Witten in six-dimensions,''
  JHEP {\bf 9803} (1998) 003
  [hep-th/9712143].
  
  %
\bibitem{Hanany:1997gh}
  A.~Hanany and A.~Zaffaroni,
  ``Branes and six-dimensional supersymmetric theories,''
  Nucl.\ Phys.\ B {\bf 529} (1998) 180
  [hep-th/9712145].
 
\bibitem{Gaiotto:2008ak} 
  D.~Gaiotto and E.~Witten,
  ``S-Duality of Boundary Conditions In N=4 Super Yang-Mills Theory,''
  Adv.\ Theor.\ Math.\ Phys.\  {\bf 13}, no. 3, 721 (2009)
  [arXiv:0807.3720 [hep-th]].


\bibitem{DHoker:2007hhe} 
  E.~D'Hoker, J.~Estes and M.~Gutperle,
  ``Exact half-BPS Type IIB interface solutions. II. Flux solutions and multi-Janus,''
  JHEP {\bf 0706}, 022 (2007)
  [arXiv:0705.0024 [hep-th]].
   E.~D'Hoker, J.~Estes, M.~Gutperle and D.~Krym,
  ``Exact Half-BPS Flux Solutions in M-theory. I: Local Solutions,''
  JHEP {\bf 0808}, 028 (2008)
  [arXiv:0806.0605 [hep-th]].


\bibitem{Assel:2011xz} 
  B.~Assel, C.~Bachas, J.~Estes and J.~Gomis,
  ``Holographic Duals of D=3 N=4 Superconformal Field Theories,''
  JHEP {\bf 1108}, 087 (2011)
  [arXiv:1106.4253 [hep-th]].
\bibitem{Bachas:2017wva}
C.~Bachas, M.~Bianchi and A.~Hanany,
``$ \mathcal{N}=2 $ moduli of AdS$_{4}$ vacua: a fine-print study,''
JHEP \textbf{08}, 100 (2018)
[arXiv:1711.06722 [hep-th]].
C.~Bachas, I.~Lavdas and B.~Le Floch,
``Marginal Deformations of 3d $N=4$ Linear Quiver Theories,''
JHEP \textbf{10}, 253 (2019)
[arXiv:1905.06297 [hep-th]].
  
  
\bibitem{Lozano:2016wrs} 
  Y.~Lozano, N.~T.~Macpherson, J.~Montero and C.~Nunez,
  ``Three-dimensional $ \mathcal{N}=4 $ linear quivers and non-Abelian T-duals,''
  JHEP {\bf 1611}, 133 (2016)
  [arXiv:1609.09061 [hep-th]].


  
\bibitem{Lozano:2015bra}
  Y.~Lozano, N.~T.~Macpherson, J.~Montero and E.~O.~Colgain,
  ``New $AdS_3 \times S^2$ T-duals with $ \mathcal{N}=\left(0,4\right) $ supersymmetry,''
  JHEP {\bf 1508} (2015) 121
  [arXiv:1507.02659 [hep-th]].

\bibitem{Kelekci:2016uqv}
  O.~Kelecki, Y.~Lozano, J.~Montero, E.~O.~Colgain and M.~Park,
  ``Large superconformal near-horizons from M-theory,''
  Phys.\ Rev.\ D {\bf 93} (2016) no.8,  086010
  [arXiv:1602.02802 [hep-th]].

  
\bibitem{Couzens:2017way}
  C.~Couzens, C.~Lawrie, D.~Martelli, S.~Schafer-Nameki and J.~M.~Wong,
  ``F-theory and AdS$_{3}$/CFT$_{2}$,''
  JHEP {\bf 1708} (2017) 043
  [arXiv:1705.04679 [hep-th]].
  
\bibitem{Macpherson:2018mif}
  N.~T.~Macpherson,
  ``Type II solutions on AdS$_{3} \times$ S$^{3} \times$ S$^{3}$ with large superconformal symmetry,''
  JHEP {\bf 1905} (2019) 089
  [arXiv:1812.10172 [hep-th]].
A.~Legramandi and N.~T.~Macpherson,
``AdS$_3$ solutions with $\mathcal{N}=(3,0)$ from S$^3\times$S$^3$ fibrations,''
[arXiv:1912.10509 [hep-th]].

 
\bibitem{Couzens:2019wls}
  C.~Couzens, H.~h.~Lam, K.~Mayer and S.~Vandoren,
  ``Black Holes and (0,4) SCFTs from Type IIB on K3,''
  arXiv:1904.05361 [hep-th].
C.~Couzens, H.~het Lam, K.~Mayer and S.~Vandoren,
``Anomalies of (0,4) SCFTs from F-theory,''
JHEP \textbf{08}, 060 (2020)
[arXiv:2006.07380 [hep-th]].


\bibitem{Lozano:2019emq}
Y.~Lozano, N.~T.~Macpherson, C.~Nunez and A.~Ramirez,
``AdS$_3$ solutions in Massive IIA with small $\mathcal{N}=(4,0)$ supersymmetry,''
JHEP \textbf{01}, 129 (2020)
[arXiv:1908.09851 [hep-th]].

 
\bibitem{Lozano:2019zvg}
Y.~Lozano, N.~T.~Macpherson, C.~Nunez and A.~Ramirez,
``Two dimensional ${\cal N}=(0,4)$ quivers dual to AdS$_3$ solutions in massive IIA,''
JHEP \textbf{01}, 140 (2020)
[arXiv:1909.10510 [hep-th]].


\bibitem{Lozano:2019jza}
Y.~Lozano, N.~T.~Macpherson, C.~Nunez and A.~Ramirez,
``1/4 BPS solutions and the AdS$_3$/CFT$_2$ correspondence,''
Phys. Rev. D \textbf{101}, no.2, 026014 (2020)
[arXiv:1909.09636 [hep-th]].

\bibitem{Lozano:2019ywa}
Y.~Lozano, N.~T.~Macpherson, C.~Nunez and A.~Ramirez,
``AdS$_3$ solutions in massive IIA, defect CFTs and T-duality,''
JHEP \textbf{12}, 013 (2019)
[arXiv:1909.11669 [hep-th]].
  %

\bibitem{Filippas:2019ihy}
K.~Filippas,
``Non-integrability on AdS$_{3}$ supergravity backgrounds,''
JHEP \textbf{02}, 027 (2020)
[arXiv:1910.12981 [hep-th]].
 
\bibitem{Speziali:2019uzn}
S.~Speziali,
``Spin 2 fluctuations in 1/4 BPS AdS$_3$/CFT$_2$,''
JHEP \textbf{03}, 079 (2020)
[arXiv:1910.14390 [hep-th]].

\bibitem{Lozano:2020bxo}
Y.~Lozano, C.~Nunez, A.~Ramirez and S.~Speziali,
``$M$-strings and AdS$_3$ solutions to M-theory with small $\mathcal{N}=(0,4)$ supersymmetry,''
JHEP \textbf{08} (2020), 118
[arXiv:2005.06561 [hep-th]].
  



\bibitem{Filippas:2020qku}
K.~Filippas,
``Holography for 2d $\mathcal{N}=(0,4)$ quantum field theory,''
[arXiv:2008.00314 [hep-th]].



\bibitem{Cvetic:2000cj}
M.~Cvetic, H.~Lu, C.~N.~Pope and J.~F.~Vazquez-Poritz,
``AdS in warped space-times,''
Phys. Rev. D \textbf{62} (2000), 122003
[arXiv:hep-th/0005246 [hep-th]].

\bibitem{Gauntlett:2006ns}
J.~P.~Gauntlett, N.~Kim and D.~Waldram,
``Supersymmetric AdS(3), AdS(2) and Bubble Solutions,''
JHEP \textbf{04} (2007), 005
[arXiv:hep-th/0612253 [hep-th]].

\bibitem{DHoker:2007mci}
E.~D'Hoker, J.~Estes and M.~Gutperle,
``Gravity duals of half-BPS Wilson loops,''
JHEP \textbf{06} (2007), 063
[arXiv:0705.1004 [hep-th]].

\bibitem{Gupta:2008ki}
R.~K.~Gupta and A.~Sen,
``Ads(3)/CFT(2) to Ads(2)/CFT(1),''
JHEP \textbf{04}, 034 (2009)
[arXiv:0806.0053 [hep-th]].

\bibitem{Chiodaroli:2009yw}
M.~Chiodaroli, M.~Gutperle and D.~Krym,
``Half-BPS Solutions locally asymptotic to AdS(3) x S**3 and interface conformal field theories,''
JHEP \textbf{02} (2010), 066
[arXiv:0910.0466 [hep-th]].
M.~Chiodaroli, E.~D'Hoker and M.~Gutperle,
``Open Worldsheets for Holographic Interfaces,''
JHEP \textbf{03}, 060 (2010)
[arXiv:0912.4679 [hep-th]].


\bibitem{Kim:2013xza}
N.~Kim,
``Comments on $AdS_2$ solutions from M2-branes on complex curves and the backreacted K\"ahler geometry,''
Eur. Phys. J. C \textbf{74}, no.2, 2778 (2014)
[arXiv:1311.7372 [hep-th]].
  
\bibitem{Corbino:2017tfl}
D.~Corbino, E.~D'Hoker and C.~F.~Uhlemann,
``AdS$_{2} \times S^{6}$ versus AdS$_{6} \times S^{2}$ in Type IIB supergravity,''
JHEP \textbf{03} (2018), 120
[arXiv:1712.04463 [hep-th]].
  
  
\bibitem{Dibitetto:2018gbk}
G.~Dibitetto and A.~Passias,
``AdS$_{2} \times S^{7}$ solutions from D0-F1-D8 intersections,''
JHEP \textbf{10}, 190 (2018)
[arXiv:1807.00555 [hep-th]].

\bibitem{Dibitetto:2018gtk}
G.~Dibitetto and N.~Petri,
``AdS$_{2}$ solutions and their massive IIA origin,''
JHEP \textbf{05} (2019), 107
[arXiv:1811.11572 [hep-th]].

  
\bibitem{Corbino:2018fwb}
D.~Corbino, E.~D'Hoker, J.~Kaidi and C.~F.~Uhlemann,
``Global half-BPS $AdS_2\times S^6$ solutions in Type IIB,''
JHEP \textbf{03}, 039 (2019)
[arXiv:1812.10206 [hep-th]].

\bibitem{Hong:2019wyi}
J.~Hong, N.~T.~Macpherson and L.~A.~Pando Zayas,
``Aspects of AdS$_{2}$ classification in M-theory: solutions with mesonic and baryonic charges,''
JHEP \textbf{11} (2019), 127
[arXiv:1908.08518 [hep-th]].

\bibitem{Dibitetto:2019nyz}
G.~Dibitetto, Y.~Lozano, N.~Petri and A.~Ramirez,
``Holographic description of M-branes via AdS$_{2}$,''
JHEP \textbf{04}, 037 (2020)
[arXiv:1912.09932 [hep-th]].

  
\bibitem{Lust:2020npd}
D.~Lust and D.~Tsimpis,
``AdS$_2$ Type-IIA Solutions and Scale Separation,''
[arXiv:2004.07582 [hep-th]].

\bibitem{Chen:2020mtv}
K.~Chen, M.~Gutperle and M.~Vicino,
``Holographic Line Defects in $D=4$, $N=2$ Gauged Supergravity,''
Phys. Rev. D \textbf{102}, no.2, 026025 (2020)
[arXiv:2005.03046 [hep-th]].



\bibitem{Corbino:2020lzq}
D.~Corbino,
``Warped $AdS_{2}$ and $SU(1,1|4)$ symmetry in Type IIB,''
[arXiv:2004.12613 [hep-th]].
 




\bibitem{Strominger:1998yg}
A.~Strominger,
``AdS(2) quantum gravity and string theory,''
JHEP \textbf{01}, 007 (1999)
[arXiv:hep-th/9809027 [hep-th]].

 
 
\bibitem{Hartman:2008dq}
T.~Hartman and A.~Strominger,
``Central Charge for AdS(2) Quantum Gravity,''
JHEP \textbf{04}, 026 (2009)
[arXiv:0803.3621 [hep-th]].


\bibitem{Alishahiha:2008tv}
M.~Alishahiha and F.~Ardalan,
``Central Charge for 2D Gravity on AdS(2) and AdS(2)/CFT(1) Correspondence,''
JHEP \textbf{08}, 079 (2008)
[arXiv:0805.1861 [hep-th]].


\bibitem{Cadoni:1999ja}
M.~Cadoni and S.~Mignemi,
``Asymptotic symmetries of AdS(2) and conformal group in d = 1,''
Nucl. Phys. B \textbf{557}, 165-180 (1999)
[arXiv:hep-th/9902040 [hep-th]].
 

\bibitem{Balasubramanian:2009bg}
V.~Balasubramanian, J.~de Boer, M.~Sheikh-Jabbari and J.~Simon,
``What is a chiral 2d CFT? And what does it have to do with extremal black holes?,''
JHEP \textbf{02}, 017 (2010)
[arXiv:0906.3272 [hep-th]].

\bibitem{Azeyanagi:2007bj}
T.~Azeyanagi, T.~Nishioka and T.~Takayanagi,
``Near Extremal Black Hole Entropy as Entanglement Entropy via AdS(2)/CFT(1),''
Phys. Rev. D \textbf{77}, 064005 (2008)
[arXiv:0710.2956 [hep-th]].
  
\bibitem{Castro:2014ima}
A.~Castro and W.~Song,
``Comments on AdS$_2$ Gravity,''
[arXiv:1411.1948 [hep-th]].


\bibitem{Cvetic:2016eiv}
M.~Cvetic and I.~Papadimitriou,
``AdS$_{2}$ holographic dictionary,''
JHEP \textbf{12}, 008 (2016)
[arXiv:1608.07018 [hep-th]].



\bibitem{Anninos:2017hhn}
D.~Anninos and D.~M.~Hofman,
``Infrared Realization of dS$_2$ in AdS$_2$,''
Class. Quant. Grav. \textbf{35}, no.8, 085003 (2018)
[arXiv:1703.04622 [hep-th]].
D.~Anninos, D.~M.~Hofman and J.~Kruthoff,
``Charged Quantum Fields in AdS$_2$,''
SciPost Phys. \textbf{7}, no.4, 054 (2019)
[arXiv:1906.00924 [hep-th]].


 

\bibitem{vanBeest:2020vlv}
M.~van Beest, S.~Cizel, S.~Schafer-Nameki and J.~Sparks,
``$\mathcal{I}$/$c$-Extremization in M/F-Duality,''
SciPost Phys. \textbf{9}, no.3, 029 (2020)
[arXiv:2004.04020 [hep-th]].

\bibitem{Mirfendereski:2020rrk}
D.~Mirfendereski, J.~Raeymaekers and D.~Van Den Bleeken,
``Superconformal mechanics of AdS$_2$ D-brane boundstates,''
[arXiv:2009.07107 [hep-th]].

\bibitem{Aniceto:2020saj}
P.~Aniceto, G.~Lopes Cardoso and S.~Nampuri,
``$R^2$ corrected AdS$_2$ holography,''
[arXiv:2010.08761 [hep-th]].


\bibitem{Maldacena:1998uz}
J.~M.~Maldacena, J.~Michelson and A.~Strominger,
``Anti-de Sitter fragmentation,''
JHEP \textbf{02}, 011 (1999)
[arXiv:hep-th/9812073 [hep-th]].
  
\bibitem{Denef:2007yt}
F.~Denef, D.~Gaiotto, A.~Strominger, D.~Van den Bleeken and X.~Yin,
``Black Hole Deconstruction,''
JHEP \textbf{03} (2012), 071
[arXiv:hep-th/0703252 [hep-th]].

\bibitem{Maldacena:2016hyu}
J.~Maldacena and D.~Stanford,
``Remarks on the Sachdev-Ye-Kitaev model,''
Phys. Rev. D \textbf{94} (2016) no.10, 106002
[arXiv:1604.07818 [hep-th]].

\bibitem{Maldacena:2016upp}
J.~Maldacena, D.~Stanford and Z.~Yang,
``Conformal symmetry and its breaking in two dimensional Nearly Anti-de-Sitter space,''
PTEP \textbf{2016} (2016) no.12, 12C104
[arXiv:1606.01857 [hep-th]].

\bibitem{Harlow:2018tqv}
D.~Harlow and D.~Jafferis,
``The Factorization Problem in Jackiw-Teitelboim Gravity,''
JHEP \textbf{02} (2020), 177
[arXiv:1804.01081 [hep-th]].

\bibitem{Hanany:1996ie}
A.~Hanany and E.~Witten,
``Type IIB superstrings, BPS monopoles, and three-dimensional gauge dynamics,''
Nucl. Phys. B \textbf{492}, 152-190 (1997)
[arXiv:hep-th/9611230 [hep-th]].

 
\bibitem{Tong:2014yna}
  D.~Tong,
  ``The holographic dual of $AdS_{3} \times  S^{3} \times S^{3} \times S^{1}$,''
  JHEP {\bf 1404} (2014) 193
  [arXiv:1402.5135 [hep-th]].

  
\bibitem{Putrov:2015jpa} 
  P.~Putrov, J.~Song and W.~Yan,
  ``(0,4) dualities,''
  JHEP {\bf 1603}, 185 (2016)
  [arXiv:1505.07110 [hep-th]].
  
  
\bibitem{Franco:2015tna} 
  S.~Franco, D.~Ghim, S.~Lee, R.~K.~Seong and D.~Yokoyama,
  ``2d (0,2) Quiver Gauge Theories and D-Branes,''
  JHEP {\bf 1509}, 072 (2015)
  [arXiv:1506.03818 [hep-th]].



\bibitem{Witten:1993yc}
E.~Witten,
``Phases of N=2 theories in two-dimensions,''
AMS/IP Stud. Adv. Math. \textbf{1}, 143-211 (1996)
[arXiv:hep-th/9301042 [hep-th]].





\bibitem{Assel:2018rcw}
B.~Assel and A.~Sciarappa,
``Wilson loops in 5d $\mathcal{N}=1$ theories and S-duality,''
JHEP \textbf{10} (2018), 082
[arXiv:1806.09636 [hep-th]].

\bibitem{Assel:2019iae}
B.~Assel and A.~Sciarappa,
``On monopole bubbling contributions to 't Hooft loops,''
JHEP \textbf{05}, 180 (2019)
[arXiv:1903.00376 [hep-th]].


\bibitem{Lozano:2020sae}
Y.~Lozano, C.~Nunez, A.~Ramirez and S.~Speziali,
[arXiv:2011.13932 [hep-th]].

\bibitem{Lozano:2021rmk}
Y.~Lozano, C.~Nunez and A.~Ramirez,
[arXiv:2101.04682 [hep-th]].



  
\bibitem{Klebanov:2007ws} 
  I.~R.~Klebanov, D.~Kutasov and A.~Murugan,
  ``Entanglement as a probe of confinement,''
  Nucl.\ Phys.\ B {\bf 796}, 274 (2008)
  [arXiv:0709.2140 [hep-th]].
  
\bibitem{Macpherson:2014eza} 
  N.~T.~Macpherson, C.~Nunez, L.~A.~Pando Zayas, V.~G.~J.~Rodgers and C.~A.~Whiting,
  ``Type IIB supergravity solutions with AdS$_{5}$ from Abelian and non-Abelian T dualities,''
  JHEP {\bf 1502}, 040 (2015)
  [arXiv:1410.2650 [hep-th]].
  Y.~Bea, J.~D.~Edelstein, G.~Itsios, K.~S.~Kooner, C.~Nunez, D.~Schofield and J.~A.~Sierra-Garcia,
  ``Compactifications of the Klebanov-Witten CFT and new AdS$_{3}$ backgrounds,''
  JHEP {\bf 1505}, 062 (2015)
  [arXiv:1503.07527 [hep-th]].
  
  
  
  
 


\bibitem{Denef:2002ru}
F.~Denef,
``Quantum quivers and Hall / hole halos,''
JHEP \textbf{10}, 023 (2002)
[arXiv:hep-th/0206072 [hep-th]].

\bibitem{Ohta:2014ria}
K.~Ohta and Y.~Sasai,
``Exact Results in Quiver Quantum Mechanics and BPS Bound State Counting,''
JHEP \textbf{11}, 123 (2014)
[arXiv:1408.0582 [hep-th]].

\bibitem{Cordova:2014oxa}
C.~Cordova and S.~H.~Shao,
``An Index Formula for Supersymmetric Quantum Mechanics,''
[arXiv:1406.7853 [hep-th]].
C.~Cordova and S.~H.~Shao,
``Counting Trees in Supersymmetric Quantum Mechanics,''
[arXiv:1502.08050 [hep-th]].




 
\bibitem{Elitzur:1985xj}
S.~Elitzur, Y.~Frishman, E.~Rabinovici and A.~Schwimmer,
``Origins of Global Anomalies in Quantum Mechanics,''
Nucl. Phys. B \textbf{273}, 93-108 (1986)





\bibitem{Aguilera-Damia:2017znn}
J.~Aguilera-Damia, D.~H.~Correa, F.~Fucito, V.~I.~Giraldo-Rivera, J.~F.~Morales and L.~A.~Pando Zayas,
``Strings in Bubbling Geometries and Dual Wilson Loop Correlators,''
JHEP \textbf{12}, 109 (2017)
[arXiv:1709.03569 [hep-th]].











\bibitem{Couzens:2018wnk}
C.~Couzens, J.~P.~Gauntlett, D.~Martelli and J.~Sparks,
``A geometric dual of $c$-extremization,''
JHEP \textbf{01}, 212 (2019)
[arXiv:1810.11026 [hep-th]].
J.~P.~Gauntlett, D.~Martelli and J.~Sparks,
``Toric geometry and the dual of $c$-extremization,''
JHEP \textbf{01}, 204 (2019)
[arXiv:1812.05597 [hep-th]].

\bibitem{Hosseini:2019ddy}
S.~M.~Hosseini and A.~Zaffaroni,
``Geometry of $\mathcal{I}$-extremization and black holes microstates,''
JHEP \textbf{07}, 174 (2019)
[arXiv:1904.04269 [hep-th]].
J.~P.~Gauntlett, D.~Martelli and J.~Sparks,
``Toric geometry and the dual of ${\cal I}$-extremization,''
JHEP \textbf{06}, 140 (2019)
[arXiv:1904.04282 [hep-th]].
H.~Kim and N.~Kim,
``Black holes with baryonic charge and $\mathcal{I}$-extremization,''
JHEP \textbf{11} (2019), 050
[arXiv:1904.05344 [hep-th]].

\bibitem{GonzalezLezcano:2020yeb}
A.~Gonzalez Lezcano, J.~Hong, J.~T.~Liu and L.~A.~Pando Zayas,
``Sub-leading Structures in Superconformal Indices: Subdominant Saddles and Logarithmic Contributions,''
[arXiv:2007.12604 [hep-th]].

\bibitem{Faedo:2020nol}
F.~Faedo, Y.~Lozano and N.~Petri,
``Searching for surface defect CFTs within AdS$_3$,''
JHEP \textbf{11} (2020), 052
[arXiv:2007.16167 [hep-th]].


\bibitem{Dibitetto:2017klx}
G.~Dibitetto and N.~Petri,
``6d surface defects from massive type IIA,''
JHEP \textbf{01} (2018), 039
[arXiv:1707.06154 [hep-th]].

\bibitem{Dibitetto:2018iar}
G.~Dibitetto and N.~Petri,
``Surface defects in the D4 $-$ D8 brane system,''
JHEP \textbf{01} (2019), 193
[arXiv:1807.07768 [hep-th]].




\bibitem{Bena:2018bbd}
I.~Bena, P.~Heidmann and D.~Turton,
``AdS$_{2}$ holography: mind the cap,''
JHEP \textbf{12}, 028 (2018)
[arXiv:1806.02834 [hep-th]].




\bibitem{Benini:2015eyy}
F.~Benini, K.~Hristov and A.~Zaffaroni,
``Black hole microstates in AdS$_{4}$ from supersymmetric localization,''
JHEP \textbf{05} (2016), 054
[arXiv:1511.04085 [hep-th]].


\bibitem{Belin:2019rba}
A.~Belin, A.~Castro, C.~A.~Keller and B.~Muhlmann,
``The Holographic Landscape of Symmetric Product Orbifolds,''
JHEP \textbf{01}, 111 (2020)
[arXiv:1910.05342 [hep-th]].
A.~Belin, N.~Benjamin, A.~Castro, S.~M.~Harrison and C.~A.~Keller,
``$\mathcal{N}=2$ Minimal Models: A Holographic Needle in a Symmetric Orbifold Haystack,''
SciPost Phys. \textbf{8}, no.6, 084 (2020)
[arXiv:2002.07819 [hep-th]].
M.~Suh,
``Uplifting supersymmetric $AdS_6$ black holes to type II supergravity,''
[arXiv:1908.09846 [hep-th]].
  
  
\bibitem{Fedoruk:2015lza}
S.~Fedoruk and E.~Ivanov,
``New realizations of the supergroup D(2, 1; \ensuremath{\alpha}) in $ \mathcal{N}=4 $ superconformal mechanics,''
JHEP \textbf{10}, 087 (2015)
[arXiv:1507.08584 [hep-th]].
I.~E.~Cunha, N.~L.~Holanda and F.~Toppan,
``From worldline to quantum superconformal mechanics with and without oscillatorial terms: $D(2,1;\alpha)$ and $sl(2|1)$ models,''
Phys. Rev. D \textbf{96}, no.6, 065014 (2017)
[arXiv:1610.07205 [hep-th]].

\bibitem{Anninos:2013nra}
D.~Anninos, T.~Anous, P.~de Lange and G.~Konstantinidis,
``Conformal quivers and melting molecules,''
JHEP \textbf{03}, 066 (2015)
[arXiv:1310.7929 [hep-th]].
D.~Anninos, T.~Anous and F.~Denef,
``Disordered Quivers and Cold Horizons,''
JHEP \textbf{12}, 071 (2016)
[arXiv:1603.00453 [hep-th]].


  
  






\bibitem{Witten:1998xy}
E.~Witten,
``Baryons and branes in anti-de Sitter space,''
JHEP \textbf{07}, 006 (1998)
[arXiv:hep-th/9805112 [hep-th]].





\end{thebibliography}
\end{document}